\documentclass{aa}  

\usepackage{graphicx}
\usepackage{txfonts}
\newcommand{\lya}{\mathrm{Ly}\ensuremath{\alpha}}
\newcommand{\civ}{C\,IV}
\newcommand{\mgii}{Mg\,II}
\newcommand{\cii}{[C\,II]}
\newcommand{\co}{CO(6-5)}
\newcommand{\cp}{C$^{+}$}

\usepackage{hyperref}

\begin{document} 

   \title{The host galaxies of radio--loud quasars at $z>$5 with ALMA}
   \titlerunning{The host galaxies of radio--loud quasars at $z>$5 with ALMA}
   \authorrunning{Mazzucchelli et al.}
   

   \author{C. Mazzucchelli\inst{1}\thanks{email: chiara.mazzucchelli@mail.udp.cl},
          R. Decarli\inst{2}, 
          S. Belladitta\inst{2,3}, 
          E. Ba\~nados\inst{3}, 
          R. A. Meyer\inst{4}, 
          T. Connor\inst{5}, 
          E. Momjian\inst{6},  
          S. Rojas-Ruiz\inst{7},  
          A.-C. Eilers\inst{8}, 
          Y. Khusanova\inst{3}, 
          E. P. Farina\inst{9}, 
          A. B. Drake\inst{10}, 
          F. Walter\inst{3}, 
          F. Wang\inst{11}, 
          M. Onoue\inst{12,13,14} 
          and 
          B. P. Venemans\inst{15} 
          }
          
    \institute{
    $^{1}$ Instituto de Estudios Astrof\'{\i}sicos, Facultad de Ingenier\'{\i}a y Ciencias, Universidad Diego Portales, Avenida Ejercito Libertador 441, Santiago, Chile.\\
    $^{2}$ INAF - Osservatorio di Astrofisica e Scienza dello Spazio di Bologna, Via Gobetti 93/3, I-40129 Bologna, Italy\\
    $^{3}$ Max Planck Institut f\"ur Astronomie, K\"onigstuhl 17, D-69117, Heidelberg, Germany\\
    $^{4}$ Department of Astronomy, University of Geneva, Chemin Pegasi 51, 1290 Versoix, Switzerland\\
    $^{5}$ Center for Astrophysics $\vert$\ Harvard\ \&\ Smithsonian, 60 Garden St., Cambridge, MA 02138, USA\\
    $^{6}$ National Radio Astronomy Observatory, P.O. Box O, 1011 Lopezville Road, Socorro, NM 87801, USA\\
    $^{7}$ Department of Physics and Astronomy, University of California, Los Angeles, 430 Portola Plaza, Los Angeles, CA 90095, USA\\
    $^{8}$ MIT Kavli Institute for Astrophysics and Space Research, 77 Massachusetts Ave., Cambridge, MA 02139, USA\\
    $^{9}$ Gemini Observatory, NSF’s NOIRLab, 670 N A’ohoku Place, Hilo, Hawai'i 96720, USA\\
    $^{10}$ Centre for Astrophysics Research, University of Hertfordshire, Hatfield, Hertfordshire AL10 9AB, U.K \\
    $^{11}$ Steward Observatory, University of Arizona, 933 N Cherry Avenue, Tucson, AZ 85721, US \\
    $^{12}$ Kavli Institute for the Physics and Mathematics of the Universe (Kavli IPMU, WPI), The University of Tokyo Institutes for Advanced Study, The University of Tokyo, Kashiwa, Chiba 277-8583, Japan \\
    $^{13}$ Center for Data-Driven Discovery, Kavli IPMU (WPI), UTIAS, The University of Tokyo, Kashiwa, Chiba 277-8583, Japan \\
    $^{14}$ Kavli Institute for Astronomy and Astrophysics, Peking University, Beijing 100871, China \\
    $^{15}$Leiden Observatory, Leiden University, PO Box 9513, 2300 RA Leiden, The Netherlands \\
    }


 
  \abstract
  {
  The interaction between radio-jets and quasar host galaxies plays a paramount role in quasar/galaxy co-evolution. However, very little has been known so far about this interaction at very high$-z$.
  Here, we present new Atacama Large Millimeter/submillimeter Array (ALMA) observations in Band 7 and Band 3 of six radio-loud quasars' host galaxies at $z > 5$. We recover \cii\,158 $\mu$m line and underlying dust continuum emission at $>2\sigma$ for five sources, while we obtain upper limits for the \co\, emission line and continuum for the remaining source. 
  At the spatial resolution of our observations ($\sim$1\farcs0-1\farcs4), we do not recover perturbed/extended morphologies or kinematics, signatures of potential mergers.
  These galaxies already host large quantities of gas ($\sim10^{10}$ M$_{\odot}$), with \cii\, luminosities of $L_{\rm \cii}\sim10^{8-9} L_{\odot}$ and \cii-based star formation rates of $30-400 M_{\odot} $yr$^{-1}$.
  Building their radio/sub-mm spectral energy distributions (SEDs), we find that in at least four cases the 1mm continuum intensity arises from a combination of synchrotron and dust emission, with an initial estimation of synchrotron contribution at 300 GHz of $\gtrsim$10\%.
  Assuming the case in which the continuum emission is due to only cold dust as upper limit, we obtain infrared (IR) luminosities of $L_{\rm IR}\sim10^{11-12} L_{\odot}$.
  We compare the properties of the sources inspected here with a large collection of radio-quiet sources from the literature, as well as a sample of radio-loud quasars from previous studies, at comparable redshift.
  We recover a potential mild decrease in $L_{\rm \cii}$ for the radio-loud sources, which might be due to a suppression of the cool gas emission due to the radio-jets. 
  We do not find any \cii\,emitting companion galaxy candidate around the five radio-loud quasars observed in Band 7: given the depth of our dataset, this result is still consistent with that observed around radio-quiet quasars.
  Further higher-spatial resolution observations, over a larger frequency range, of high$-z$ radio-loud quasars hosts will allow for a better understanding of the physics of such sources.
  }

   \maketitle
%

\section{Introduction}
Quasars are among the most luminous, non-transient sources in the universe, and they can be observed at very large distances, hence at large look-back time. In the last two decades, the number of quasars known at high redshift ($z\gtrsim 5$, i.e. within one Gyr from the Big Bang) increased dramatically, with more than 400 sources discovered so far (e.g. \citealt{fan2001},\citeyear{fan2006}, \citealt{banados2016}, \citealt{mazzucchelli2017b}, \citealt{matsuoka2018}, \citealt{reed2019}, \citealt{yang2019}, \citeyear{yang2023}, \citealt{wenzl2021}) up to $z\sim7.5$ (\citealt{banados2018a}, \citealt{yang2020}, \citealt{wangf2021}; see \citealt{fan2023} for a recent review). Very large supermassive black holes (SMBH) have been observed in their centers ($>10^{8} M_{\odot}$; e.g.\,\citealt{willott2015}, \citealt{yang2021}, \citealt{farina2022}, \citealt{mazzucchelli2023}), surrounded by gas already enriched at (super-)solar metallicities (e.g.\,\citealt{xu2018}, \citealt{onoue2020}, \citealt{lai2022}).

Observations of the stellar radiation from the host galaxies of high-redshift quasars have been challenging (e.g.\,\citealt{decarli2012}, \citealt{mechtley2012}, \citealt{marshall2020}), due to the overwhelming radiation from the nuclear engine. Only very recently, observations with the \textit{James Webb Space Telescope} (JWST) have uncovered such stellar emission in a few sources (e.g.\,\citealt{ding2023}, \citealt{marshall2023}, \citealt{stone2023a}, \citeyear{stone2024}, \citealt{yue2024}, \citealt{onoue2024}). Conversely, the properties of the host galaxies can be more easily explored with observations of the cool gas and dust emission from the interstellar medium (ISM), recovered in the (sub-)mm regime at high$-z$. In particular, the \cii\,158 $\mu$m emission line is the main coolant of the ISM, and can emit up to $3\%$ of the total galactic output (e.g.\,\citealt{carilli2013}). Initial studies of a few, massive sources known at that time were pursed with the IRAM Plateau de Bure interferometer in the 2010s (e.g.\,\citealt{maiolino2005}, \citealt{walter2009}, \citealt{wang2010}). Afterwards, the advent of the Atacama Large Millimeter/submillimeter Array (ALMA) paved the way for the exploration of large, statistically significant samples of quasars (e.g.\,\citealt{wang2013}, \citealt{willott2015},\citeyear{willott2017}, \citealt{trakhtenbrot2017}, \citealt{decarli2018}, \citealt{venemans2018},\citeyear{venemans2020}, \citealt{izumi2018},\citeyear{izumi2019},\citeyear{izumi2021}, \citealt{nguyen2020}, \citealt{wangf2024}). So far, almost 80 quasars' host galaxies have been observed in the (sub-)mm at $z>5$. The main results of these observations are: $a)$ quasars' hosts already contain large reservoirs of gas and dust ($M_{\rm dust} > 10^{8} M_{\odot}$), and are very rapidly forming stars, with star formation rates (SFR) up to $>100-1000 M_{\odot}$ yr$^{-1}$. $b)$ the central SMBHs seem to be overmassive with respect to predictions from the SMBH-host galaxy mass relation observed in the local universe (e.g.\,\citealt{venemans2016}, \citealt{pensabene2020}, \citealt{neeleman2021}), although this discrepancy seems less marked at lower luminosities/black hole masses (e.g.\,\citealt{izumi2021}). $c)$ quasars' hosts presents a variety of kinematics--e.g.\,rotation supported, dispersion dominated, or disturbed by close companions and/or mergers events (e.g.\,\citealt{neeleman2021})--while observational studies provide mixed results for the presence of major molecular/cool gas outflows (e.g.\,\citealt{decarli2018}, \citealt{bischetti2019}, \citealt{novak2020}, \citealt{meyer2022a}). $d)$ $\gtrsim 20 - 40 \%$ of these sources are surrounded by overdensities of galaxies on tens of kpcs scales (e.g.\citealt{decarli2017},\citealt{trakhtenbrot2017},\citealt{nguyen2020}, \citealt{venemans2020}), which are extremely dust/gas-rich and severely obscured (e.g.\,\citealt{mazzucchelli2019}). Despite the large sample of high-redshift quasars now observed in the sub-mm, almost all have been radio-quiet (RQ), due to the absence of dedicated ALMA follow-up of radio-loud quasars.

Indeed, a fraction of quasars ($\sim 10\%-20\%$, irrespective of redshift up to $z\gtrsim6$, i.e.\,within the Epoch of Reionization, EOR; e.g.\,\citealt{banados2015b}, \citealt{liu2021}; but see also \citealt{sbarrato2022}) show strong radio emission, associated with powerful jets. These sources are defined as radio-loud (RL) quasars, i.e. with a radio-loudness parameter $R_{4400} = f_{\nu,{\rm 5GHz}} / f_{\rm 4400 \AA} > 10$ \citep{kellerman1989} or $R_{2500} = f_{\nu,{\rm 5GHz}} / f_{\rm 2500 \AA} > 10$ \citep{jiang2007}, where $f_{\nu,{\rm 5GHz}}$, $f_{\rm 4400 \AA}$ and $f_{\rm 2500 \AA}$ are the flux densities at 5GHz, 4400 $\AA$ and 2500 $\AA$, respectively. These two quantities provide similar results for an un-obscured type-1 quasar; in this paper we will consider $R_{2500}$ as our radio-loudness parameter. Radio-jets are thought to be key elements in the host galaxy-SMBH (co-)evolution, via the so-called `radio-mode feedback'. Indeed, on one hand, jets can reduce star formation by sweeping away the gas reservoir (e.g.\,\citealt{villarmartin2014}), while on the other hand they might enhance the formation of new stars, especially at high$-z$, via shock-induced gravitational collapse of gas clouds in an in-homogeneous and dense medium (e.g.\,\citealt{silk2013}, \citealt{salome2015}, \citealt{fragile2017}). Moreover, RL quasars are routinely observed in rich galactic environments (e.g.\,\citealt{wylezalek2013}, \citealt{venemans2007}), and are considered the best targets to discover overdensities of galaxies in the EOR (e.g.\,\citealt{zheng2006}, \citealt{ajiki2006}, \citealt{bosman2020}).
Currently, there are 50 RL quasars known at $z \gtrsim 5$ (see \citealt{banados2021} and references there-in), out of which 5 are classified as blazars (i.e.\,with the relativistic jets aligned with our line of sight; e.g.\,\citealt{belladitta2019}, \citeyear{belladitta2020}, \citealt{banados2024a}); more than half of these sources were identified in the last two years (e.g.\, \citealt{gloudemans2022}, \citealt{banados2023}, \citealt{belladitta2023}). Very few of their host galaxies have been studied so far. \cite{rojas2021} observed the $z\sim5.8$ RL quasar PSO352-15 with a collection of (sub-)mm and radio data, obtained with ALMA, the NOrthern Extended Millimeter Array (NOEMA) and the upgraded Giant Metrewave Radio Telescope (uGMRT). They found that the total sub-mm emission can not be uniquely explained by dust emission, and that synchrotron radiation contributes down to these frequencies. \cite{khusanova2022} observed four other $z>6$ RL quasars in the Northern Hemisphere with NOEMA, targeting their dust and \cii\,emission lines. While the host of the highest reshift ($z\sim6.8$) RL quasar known so far is not detected, some of the remaining hosts show asymmetric line profiles--indicative of a potential merger or strong outflow--and SFRs comparable to RQ quasars hosts. Very recently, \cite{banados2024a} discovered a blazar at $z\sim7$, and \cite{banados2024b} reported NOEMA (2021) and ALMA (2022) observations of this source. They recovered a large ($>30$\%) flux variation within only 10 months, and they estimated a synchrotron contribution to the observed $\sim$230 GHz flux as high as $\sim$80\%.

Here, we present new ALMA observations of the dust continuum and \cii\,or \co\,emission lines of the host galaxies of six $5<z<6$ RL quasars, and one RQ source\footnote{At the time of observations, J2053+0047 was miss-classified as RL, see Section \ref{sec:sampobs} for more details.}. This paper is organized as follows: in Section \ref{sec:sampobs} we report our sample and observations; in Section \ref{sec:results} we describe our methods to derive continuum and emission lines measurements; in Section \ref{sec:analysis} we report our redshift estimates, present radio-mm Spectral Energy Distribution (SEDs) of our RL quasars, calculate gas and \cii\,luminosities, SFRs, dust, dynamical and molecular gas masses, and place them in the context of RQ quasars host galaxies' observations recovered from the literature. We also compare the infrared (host galaxies) and UV (nuclear emission) properties, and search for the presence of further line emitting galaxies in their fields. Finally, we discuss our results in Section \ref{sec:disc} and present our conclusions in Section \ref{sec:conc}. 

Throughout this work, we use a cosmology, with $H_{0} = 67.8$ km s${-1}$ Mpc$^{-1}$, $\Omega_{M} = 0.3$ and $\Omega_{\Lambda} = 0.7$. The age of the universe at $z=5.5$ is 1.04 Gyr, and the transverse physical scale is 6.1 kpc per arcsec. All magnitudes are reported in the AB system. 
\section{Sample and Observations} \label{sec:sampobs}
For this study, we focused on high-redshift ($z>5$) RL quasars, observable from ALMA (Declination $<20$ deg). At the time of the submission of the proposal (April 2018), the source J2053+0047 was tentatively classified as RL by \cite{banados2015a}, based on data from the Faint Images of the Radio Sky at Twenty cm survey (FIRST, \citealt{becker1995}), with a radio-loudness parameter of $R_{4400}= 44.1 \pm 18.7$. However, \cite{liu2021} showed it to be RQ, based on deeper follow-up data with the Karl G.\,Janski Very Large Array (VLA), with $R_{4400} < 9.7$. We still provide our measurements and results for this quasar in the paper, but label it as RQ. We list in Table \ref{tab:sample1} the general properties of our sample, i.e.\,coordinates, literature redshifts, magnitude at rest-frame 1450 \AA,\,and radio-loudness parameter.

ALMA Cycle 7 observations (program ID: 2019.1.00840.S, PI: Mazzucchelli) were obtained between October and December 2019. For 5 sources, we targeted the \cii\,emission line, falling in Band 7 at 5$<\,z\,<$5.8. The redshift of two sources ($z\sim5.9$; J2228+0110 and J2053+0047) places their \cii\,line close to the gap between Band 6 and 7, hence, in these cases, we targeted the \co\,emission line, recovered in Band 3. We chose a compact configuration C(43-2) to avoid flux losses, with 44 antennas used. We centered two spectral windows, slightly overlapping, on the expected observed frequency of the \cii\,or \co\,emission line. The remaining two spectral windows were placed to recover the continuum emission at slightly different frequencies. A bandwidth of 1.875 GHz was covered per spectral window. The data were reduced with CASA (McMullin et al. 2007) pipeline for ALMA (version 6.2.1-7), with standard configuration. The cubes were imaged using the CASA task \texttt{tclean} with weighting and robust parameters set to 'briggs' and '2', respectively, to maximize the signal-to-noise ratio (SNR), following similar works from the literature \citep{decarli2018}. We produce cubes with a spectral resolution width of 30 km s$^{-1}.$ The typical rms noise per 30 km s$^{-1}$ channel is $\sim$0.50 mJy beam$^{-1}$, with a typical synthesized beam size of 1\farcs35$\times$1\farcs10.
%
%
\begin{table*}
\label{tab:sample}
\caption{Sample of quasars in this study.}    
\label{tab:sample1}      
\centering          
\begin{tabular}{l c c c c c c c}     
\hline\hline       
Name & R.A.    & Decl.   & $z$ & $z$ method & $M_{1450}$ & R$_{2500}$ & Reference \\
     & (J2000) & (J2000) &     &           &            &            & Disc / $z$ / $M_{1450}$\\
\hline                    
   J0131-0321$^{\dagger}$ & 01:31:27.3400 & --03:21:00.180 & 5.1904 $\pm$ 0.0003 & \mgii & $-$29.11 & 116 $\pm$ 9 & 1 / 2 / 1\\
   PSO055-00 & 03:41:41.8580 & --00:48:12.740 & 5.68 $\pm$ 0.05 & template & $-$26.37 & 83 $\pm$ 9 & 3 / 3 / 4\\
   PSO135+16 & 09:01:32.6530 & 16:15:06.830 & 5.63 $\pm$ 0.05 & template & $-$25.91 & 177 $\pm$ 18 & 3 / 3 / 4\\
   J1034+2033 & 10:34:18.6500 & 20:33:00.220 & 5.01 $\pm$ 0.02 & z\_VI$^{\ddagger}$ & $-$27.76 & 47 $\pm$ 4 & 5 / 6 / 7\\
   J2228+0110 & 22:28:43.5350 & 01:10:32.200 & 5.903 $\pm$ 0.0002 & $\lya$ & $-$24.50 & 71 $\pm$ 15 & 8 / 9 / 8\\  
   PSO352--15 & 23:29:36.8362 & --15:20:14.460 & 5.84 $\pm$ 0.02 & template & $-$25.60 & 1470 $\pm$ 110 & 10 / 10 / 10\\
\hline               
   J2053+0047$^{\star}$ & 20:53:21.766 & +00:47:06.80 & 5.92 $\pm$ 0.03 & $\lya$ & $-$25.30 & $<$9.7$^{\mp}$ & 11 / 11 / 11 \\
\hline
\end{tabular}
\tablefoot{The radio-loudness parameter is from \citeauthor{banados2021} (\citeyear{banados2021}; and references therein), with the exception of POS352-15, where $R_{2500}$ is from \cite{rojas2021}. Further references are: 1) \cite{yi2014}; 2) this work; 3) \cite{banados2015a}; 4) \cite{banados2016}; 5) \cite{schneider2010} ; 6) \cite{paris2018} ; 7) \cite{banados2021}; 8) \cite{zeimann2011}; 9) \cite{roche2014}; 10) \cite{banados2018c}; 11) \cite{jiang2009}.
           $^{\dagger}$Classified as blazar (see Section \ref{sec:sampobs} for details).
           $^{\ddagger}$ See \cite{paris2018} for a discussion on different redshift estimates.
           $^{\star}$ We note J2053+0047 was reclassified as a RQ quasar with deeper radio VLA observations (see Section \ref{sec:sampobs} for details).
           $^{\mp}$ Limit on radio-loudness parameter is obtained from \cite{liu2021}}
\end{table*}
%
%

Below we report notes on individual objects in the sample and further details on their observations.
\subsection{J0131--0321}
\cite{yi2014} discovered this very luminous source ($i$ = 18.47) by cross-matching the Sloan Digital Sky Survey (SDSS) Data Release 7 \citep{schneider2010} and the Wide-Field Infrared Survey Explorer (WISE; \citealt{wright2010}) catalogs, with further confirmation via spectroscopic observations at the Lijiang 2.4m telescope, and with MaGE and FIRE at the Magellan telescope. By fitting the \mgii\, emission line and underlying continuum emission, the authors estimated a redshift of $z=5.18 \pm 0.02$, a large black hole mass of 2.7$\times$10$^{9} M_{\odot}$ and a bolometric luminosity of $L_{\rm bol} \sim 10^{48}$ erg s$^{-1}$, with a resulting super-Eddington rate of $\sim$3.14. Given a radio flux density of 33 mJy at 1.4 GHz, this source presented a large radio-loudness parameter of $\sim$100. \cite{ghisellini2015} highlighted the potential blazar nature of J0131--0321 with $Swift$-XRT X-ray observations. Fitting its SED with an accretion disk+torus+jet model, they estimated a viewing angle of 3-5 deg, and a black hole mass of $\sim 10^{10} M_{\odot}$ (with a factor of $\gtrsim$2 of systematic uncertainty), allowing for a sub-Eddington accretion rate of $\sim$0.8. High-resolution Very Large Baseline Interferometer (VLBI) observations with the European VLBI Network (EVN) at 1.7 GHz, obtained by \cite{gabanyi2015}, showed a compact source with high brightness temperature ($T_{\rm B} < 10^{11}$ K) and moderate Doppler boosting factor ($\gtrsim$6). Moreover, they noted a strong radio flux variability with respect to the FIRST survey, supporting the blazar hypothesis. Finally, this source was observed with the uGMRT at 323 MHz, but radio frequency interference contaminated the data, preventing any measurement \citep{shao2020}. Nevertheless, using observations at 150 MHz from the TIFR GMRT Sky Survey (TGSS; \citealt{itema2017}), the authors could estimate a radio power law slope of $\alpha_{\nu} \sim 0.3$ (with $S_{\nu} \propto \nu^{\alpha}$).

J0131--0321 was observed with ALMA on 2019 Nov 28, with a total on-source exposure time of 726 s. The synthesized beam size is 1\farcs06$\times$\,0\farcs94, with a position angle (PA) of 82 deg. The rms noise on 30 km s$^{-1}$ channels is 0.49 mJy\,beam$^{-1}$.
\subsection{PSO055--00}
This quasar was selected by mining the Pan-STARRS and FIRST surveys, and confirmed by spectroscopic follow-up with Palomar/DBSP and VLT/FORS2 \citep{banados2015a}. A reduced $\chi^{2}$ fit of the observed spectrum to a composite high$-z$ quasar template yielded a redshift of $z=5.68 \pm 0.05$. PSO055--00 presents a radio flux density at 1.4GHz of $2.14 \pm 0.14$ mJy, as obtained from the FIRST Survey, and a radio-loudness parameter $R_{2500} = 83 \pm 9$ \citep{banados2021}.

Our ALMA observations of this quasar were obtained on 2019 Dec 17 with a total on-source exposure time of 816 s. The measured rms noise is 0.54 mJy\,beam$^{-1}$ on 30 km s$^{-1}$ channels, while the beam size is 1\farcs39\,$\times$\,1\farcs05, with PA=81 deg.
\subsection{PSO135+16}
PSO135+16 was selected as a high$-z$ quasar candidate in a similar fashion as PSO055--00. It was then confirmed spectroscopically with LBT/MODS and VLT/FORS2 \citep{banados2015a}. It is located at $z = 5.68 \pm 0.05$, as measured by fitting its spectrum to a reference template. Its radio-loudness parameter is $R_{2500} = 177 \pm 18$, with a radio flux density of $f_{\rm 1.4GHz} = 3.04\pm0.15$ mJy, as obtained from the FIRST survey. 

ALMA observed PSO135+16 on 2019 Dec 02 with exposure time of 1058 s. The beam size is of 1\farcs39\,$\times$\,1\farcs21, PA=$-$30 deg, and the rms noise is of 0.50 mJy\,beam$^{-1}$ on 30 km s$^{-1}$ channels.
\subsection{J1034+2033}
This bright ($z=19.79$) source was firstly presented as part of the SDSS-DR7 quasar catalog \citep{schneider2010}. It is located at redshift $z=5.01$\footnote{We considered here the redshift from visual inspection and as uncertainty the corresponding dispersion of Z\_VI - Z\_MGII. All the other redshift measurements in this case are consistent with the reported value within the uncertainty.} \citep{paris2018}. J1034+2033 has a radio flux density $f_{\rm 1.4GHz} = 3.96 \pm 0.15$ mJy, from the FIRST survey, and a radio-loudness parameter $R_{2500} = 47 \pm 4$. \cite{shao2020} observed J1034+2033 with the uGMRT, measuring a flux density of 2.97$\pm$0.19 mJy at 323 MHz, and fitting a radio spectral slope of $\alpha_{\nu} \sim 0.18$. 

Our ALMA data for this quasar were obtained on 2019 Dec 04, with a rms noise of 0.78 mJy\,beam$^{-1}$ on 30 km s$^{-1}$ channels, and a beam size of 1\farcs36\,$\times$\,1\farcs04, PA=$-$33 deg. The total exposure time was 1058s.
\subsection{J2228+0110}
\cite{zeimann2011} selected J2228+0110 as a quasar candidate by cross-matching the Stripe82 SDSS and Stripe 82 VLA surveys. Its high$-z$ quasar nature was confirmed by observations with the Keck/LRIS spectrograph. J2228+0110 is optically faint (with an observed magnitude $z=22.3$), with a radio flux density of $f_{\rm 1.4GHz} = 0.31 \pm 0.06$ mJy and a radio-loudness parameter of $R_{2500} \sim 70$. A tentative detection of a Lyman Alpha Halo (LAH) around this quasar was firstly presented by \cite{roche2014}, using GTC-OSIRIS long-slit spectroscopy. Deep (11h) MUSE observations confirmed the presence of an extended ($\sim$6\farcs0, i.e. $\sim$ 30 pkpc at $z\sim6$) and bright ($\sim 2 \times 10^{-16}$ erg s$^{-1}$ cm$^{-2}$) LAH towards the south-east direction (\citealt{drake2019}, \citealt{farina2019}).

ALMA targeted J2228+0110 on 2019 Oct 12, with an exposure time of 3810 s. The measured rms noise is 0.32 mJy\,beam$^{-1}$ on 30 km s$^{-1}$ channels, and the beam size is 1\farcs44\,$\times$\,1\farcs28, PA=$-$78 deg.
\subsection{PSO352--15}
\cite{banados2018c} selected PSO352--15 as a quasar candidate from the Pan-STARRS survey, and confirmed it with spectroscopic observations with Magellan/LDSS3. This quasar was initially located at $z = 5.84$, via template fit, and is optically faint ($z=21.22$). With very bright radio flux density measurements from available radio surveys and dedicated VLA follow-up, it is one of the radio loudest sources at $z>5$. Further high-resolution radio observations with the Very Long Baseline Array (VLBA) resolved the source into different components with a spatial extension of $\sim$1.3 kpc, and a morphology consistent with a one-sided jet or a compact symmetric object \citep{momjian2018}. \cite{rojas2021} presented a collection of multi-wavelength observations for this source, included ALMA data collected as part of this program. They re-calculated the radio-loudness parameter as $R_{2500} \sim 1350$, and showed evidence for a break in the radio spectral distribution. Intriguingly, they found signs for synchrotron emission contribution down to the rest-frame infrared regime (see also Section \ref{sec:sed} below). We still report PSO352-15 here as part of the RL sample and we re-calculate its properties for consistency (see Section \ref{sec:results}).

This object was observed on 2019 Nov 28, with an exposure time of 756 s. We measured a rms noise of 0.58 mJy\,beam$^{-1}$, and a beam size of 1\farcs27\,$\times$\,1\farcs00, PA=86 deg.
\subsection{J2053+0047}
This faint ($z = 21.41$) quasar at redshift $z = 5.92$ was selected by \cite{jiang2009} using SDSS survey data. \cite{banados2015a} identified a 3$\sigma$ detection in the FIRST survey data, with a radio flux density of $f_{\rm 1.4GHz} = 434 \pm 143 \mu$Jy, obtaining a radio-loudness parameter of $\sim$44, signaling this as a RL quasar. Deeper observations with the VLA provided new flux density limits of $f_{\rm 1.4GHz} \sim 13 \mu$Jy, yielding an upper limit on the radio-loudness parameter as low as $<$9.7 \citep{liu2021}. Hence, J2053+0047 has been currently reclassified as a RQ source. We still report the data collected for this source here for completeness.

Our ALMA observations of this source was obtained on 2019 Oct 13. The total exposure time was 2570 s, and we measure a beam size of 1\farcs56\,$\times$\,1\farcs20, PA=65 deg, and a rms noise of 0.23 mJy\,beam$^{-1}$.
\section{Results} \label{sec:results}
We describe here our steps to recover the properties of the \cii, \co\, emission lines, and the underlying dust continuum.
\subsection{[CII] line measurements} \label{sec:ciimeasure}
\begin{figure*}
   \centering
   \includegraphics[width=0.85\hsize]{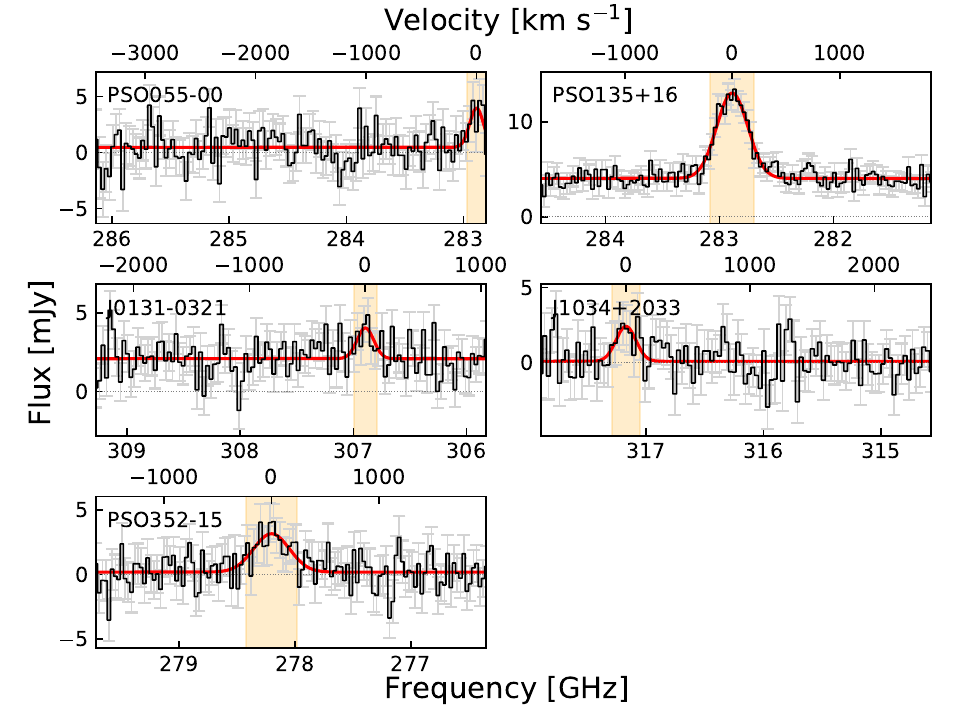}
   \caption{ALMA \cii\, emission line and underlying continuum of the quasars in our sample, re-sampled to 30 km s$^{-1}$ velocity bin, considering extraction apertures of different radii (see Section \ref{sec:ciimeasure} and Table \ref{tab:1dcii}). The error (light grey) is obtained by taking the pixels rms, and re-scaling the values considering the number of pixels in the aperture and the beam size. The continuum+Gaussian fit is plotted with a red solid line. We show the channels used to extract the continuum-subtracted \cii\,maps ($\pm$1.4$\sigma$) with yellow shading (see Section \ref{sec:ciimeasure}). }
              \label{fig:1dciiline}%
    \end{figure*}

We target the \cii\,emission line in five quasars: J0131-0321, PSO055-00, PSO135+16, J1034+2033 and PSO352-15.
In order to properly account for potentially extended sources, we extract our one dimensional spectrum for each source in an aperture centered on the quasar position (e.g.\,\citealt{novak2019}, \citealt{rojas2021}). We determine the best aperture by extracting the flux of the quasars in concentric apertures from 0\farcs1 up to 3\farcs0, and choosing the radius at which the flux encounters a plateau. Once the best aperture is found, we extract the one dimensional spectra, and fit them with a flat continuum and a Gaussian line. We detect the \cii\,emission line in all sources. The emission line for the quasar PSO055-00 is found at the edge of the bandwidth: this is due to the large uncertainty of the literature redshift derived from template fit of the rest-frame UV spectrum. Therefore, caution should be taken when considering its \cii\,emission line value, and the derived quantities. We report the 1D spectra in Figure \ref{fig:1dciiline}, and the values of the fit and apertures radii in Table \ref{tab:1dcii}.

\begin{figure*}
   \centering
   \includegraphics[width=0.85\hsize]{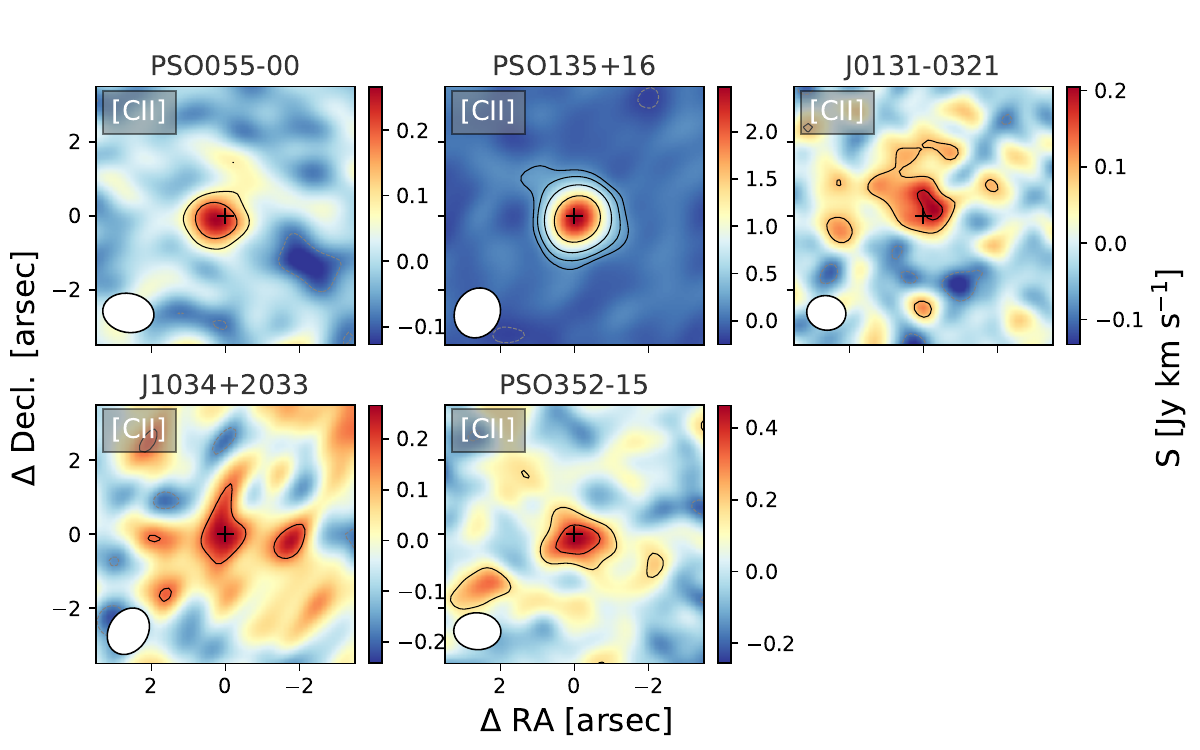}
   \caption{ALMA continuum-subtracted \cii\, maps. The continuum black/dashed grey contours highlights the $\pm$2$\sigma$,4$\sigma$,8$\sigma$..levels. The beam size is shown with a white ellipse, while the rest-frame UV/optical position of the quasar is reported with a black cross.
   The morphology of all the sources is broadly consistent with the beam size, hence we consider them as unresolved at the spatial resolution and SNR of this dataset.
   }
              \label{fig:1dciimap}%
    \end{figure*}

We create continuum-subtracted \cii\,emission line cubes in the following way. We choose the line spectral windows centered on the frequency peak of the Gaussian fit and width corresponding to $\pm$1.4$\sigma$; this recovers $\sim 83\%$ of the total line flux and maximizes the SNR, in case of a perfectly Gaussian line profile (e.g.\,\citealt{decarli2018}, \citealt{novak2020}, appendix A). The continuum subtraction is done in the \textit{uv} space using the CASA task \textit{uvcontsub}. We then re-image the continuum subtracted cubes with \texttt{tclean}, with robust=2.  We report the \cii\,continuum-subtracted moment-0 maps in Figure \ref{fig:1dciimap}.
We fit the continuum-subtracted \cii\,map with a 2D Gaussian function within CASA, selecting a small rectangular region around the source. We report the flux values and sizes of the sources in our sample in Table \ref{tab:2dfit}. We note that the maps are built considering only $83\%$ of the total line flux, therefore the 2D flux need to be corrected by a factor of $\sim$1.20 (e.g.\,\citealt{decarli2018}).
We make a first order assessment of the kinematics of these sources (see Appendix A). However, due to the low SNR and the limited spatial resolution of the maps, we are not able to draw stronger conclusions on their kinematics, for which additional deeper, and higher resolution observations are needed.
It is worth noting that our results for the quasar PSO352-15 are consistent with those presented in \cite{rojas2021}, and we direct the reader to this work for further insights in the morphological study of this source.

In the remainder of the paper, we use the \cii\,emission line flux obtained by the 2D Gaussian fit, corrected by a factor of 1.20.
%
\begin{table*}
\caption{Results from spectral fit of the sources observed in Band 7.}             
\centering          
\label{tab:1dcii}
\begin{tabular}{l c c c c c c}     
\hline\hline       
Name & $\nu_{\rm obs}$ (\cii) & $z$ (\cii) & FWHM (\cii) & $F_{\rm line}$ (\cii) & $F_{\rm cont}$  & $r_{\rm aper}$\\
     & [GHz] &  & [km s$^{-1}$] & [Jy km s$^{-1}$]  & [mJy] & [\farcs] \\
\hline                    
   J0131-0321 & 306.895$\pm$0.03 & 5.1928$\pm$0.0006 & 170$\pm$70 & 0.35$\pm$0.20 & 2.10$\pm$0.10 & 1.4\\
   PSO055-00$^{\star}$  & 283.889$\pm$0.02 & 5.7183$\pm$0.0005 & 150$\pm$60 & 0.55$\pm$0.30 & 0.46$\pm$0.20 & 2.3\\
   PSO135+16 &  282.889$\pm$0.008 & 5.7183$\pm$0.002 & 340$\pm$29 & 3.30$\pm$0.30 & 4.00$\pm$0.10 & 1.6\\
   J1034+2033 & 317.172$\pm$0.03 & 4.9921$\pm$0.0005 & 190$\pm$80 & 0.47$\pm$0.30 & 0.064$\pm$0.10 & 1.3\\
   PSO352--15 & 278.202$\pm$0.03 & 5.8315$\pm$0.0007 & 390$\pm$90 & 1.20$\pm$0.040 & 0.18$\pm$0.20 & 1.7\\
\hline
\end{tabular}
\tablefoot{The spectra were extracted from different apertures ($r_{\rm aper}$, see Section \ref{sec:ciimeasure}).
           $^{\star}$ We note that PSO055-00's line emission is recovered at the edge of the bandwith. Therefore, caution should be taken when considering its \cii\,emission line value and all derived quantities.
           }
\end{table*}
\subsection{CO(6-5) line measurements} \label{sec:co65measure}
Observations in Band 3, targeting the \co\,emission lines, were obtained for two quasars: J2228+0110 and J2053+0047. No clear emission is seen in the reduced cubes. We extract the one dimensional spectrum from the central pixel, i.e. at the nominal position of the quasar. No line nor continuum emission were detected.

\begin{figure}
   \centering
   \includegraphics[width=0.95\hsize]{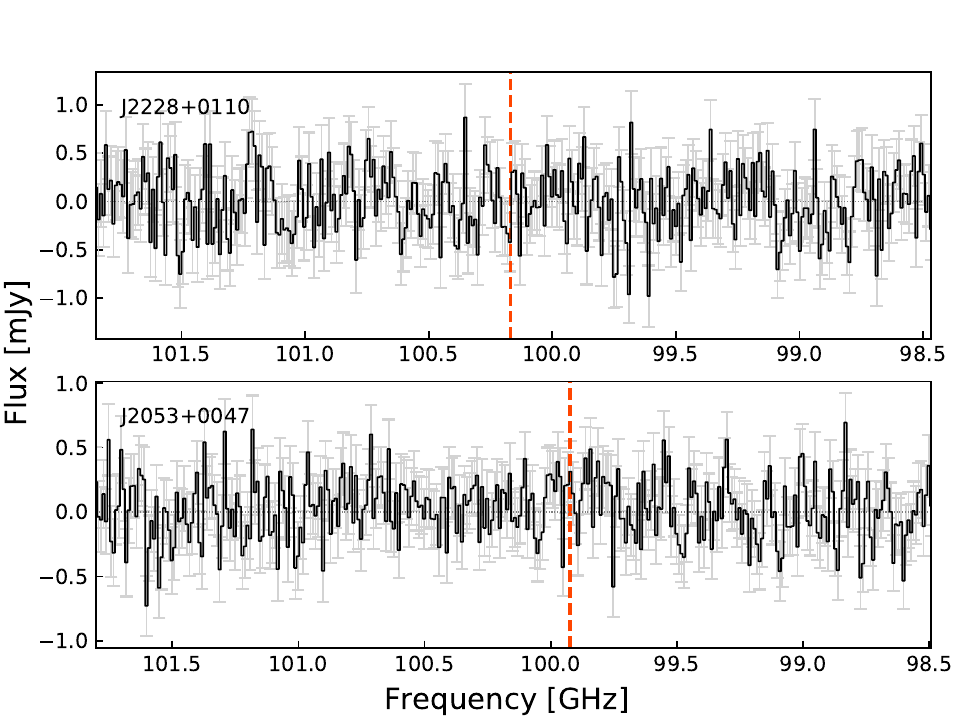}
   \caption{ALMA spectra of J2228+0110 and J2052+0047 observed in Band 3. No continuum nor \co\,emission line is detected. The expected locations of the \co\,line at the redshifts of the quasars are shown with orange dashed lines.}
              \label{fig:1dco65line}%
    \end{figure}

We can place upper limits to the continuum emission at 2$\sigma$ by considering the mean rms noise on the extracted 1D spectrum. We also estimate upper limits on the emission line flux at 3$\sigma$, and considering a typical line width of 300 km s$^{-1}$ (e.g.\,\citealt{decarli2022}). We show in Figure \ref{fig:1dco65line} the extracted one dimensional spectra, and in Table \ref{tab:co65} the continuum and \co\, emission line flux limits.
%
\begin{table}
\caption{Limits on the continuum and \co\, emission line fluxes (see Section \ref{sec:co65measure}), and derived limits on the line and infrared luminosities (see Section \ref{sec:co65analysis}).}             
\centering          
\label{tab:co65}
\begin{tabular}{l c c}     
\hline\hline       
Name  & J2228+0110 & J2053+0047 \\
\hline                    
   $F_{\rm \co}$ [Jy km s$^{-1}$] & $<$0.68 & $<$0.53 \\
   $F_{\rm cont,690  GHz}$ [mJy] & $<$0.64 & $<$0.47 \\
   $L_{\rm CO(6-5)}$ [$10^{8}\, L_{\odot}$] & $<$2.4 & $<$1.9 \\
   $L^{'}_{\rm CO(6-5)}$ [$10^{10}\, L_{\odot}$] & $<$2.2 & $<$1.8 \\
   $L_{\rm FIR}$ [$10^{12}\, L_{\odot}$] & $<$46 & $<$34 \\
   $L_{\rm TIR}$ [$10^{12}\, L_{\odot}$] & $<$67 & $<$50 \\
\hline
\end{tabular}
\end{table}
\subsection{Continuum measurements}
We recover the continuum emission at $\nu_{\rm rf} \sim 1900$ GHz and $\nu_{\rm rf} \sim 1970$ GHz in Band 7 in the lower and upper side bands, respectively. In the case of Band 3, we target the continuum at $\nu_{\rm rf} \sim 690$ GHz (lower) and $\nu_{\rm rf} \sim 620$ GHz (upper side band).

We estimate the continuum properties for the sources observed in Band 7 as follows. We create pure continuum maps for the lower and upper side bands by collapsing the frequency channels not incorporating the \cii\,emission line (see Section \ref{sec:ciimeasure}). In both cases the continuum emission is recovered in all the sources. We model each continuum emission from the maps with a 2D Gaussian function, using CASA. The measured sizes are consistent with the beam size. We produce similar continuum maps for the two sources observed in Band 3, but no emission is recovered in either case.

\begin{figure*}
   \centering
   \includegraphics[width=0.95\hsize]{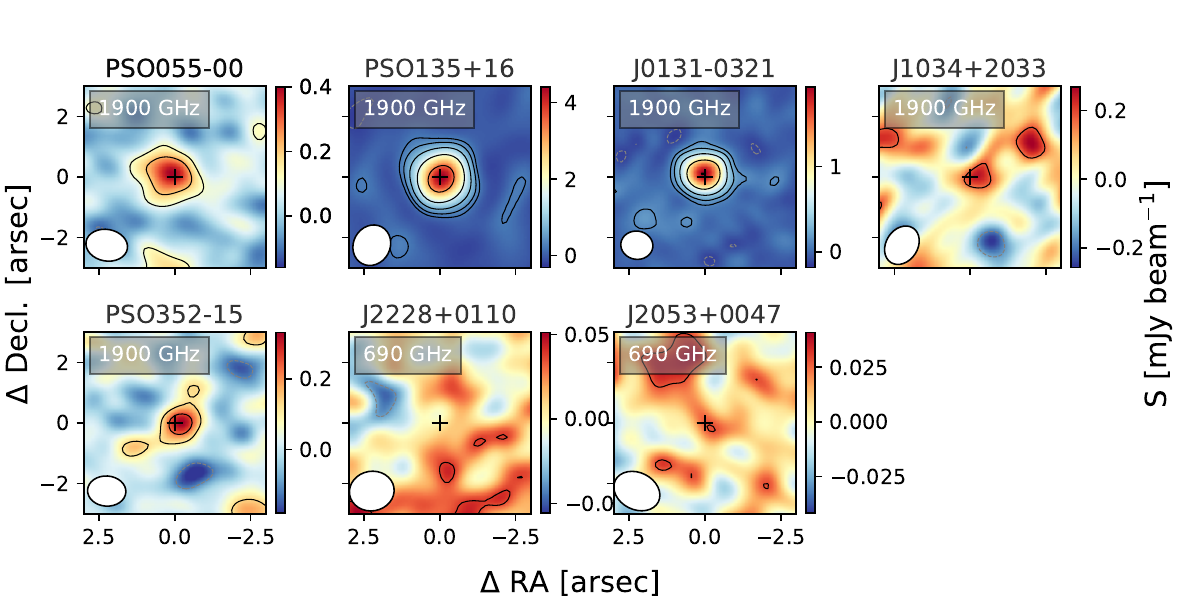}
   \caption{ALMA continuum emission maps at rest-frame 1900 GHz (Band 7) and 690 GHz (Band 3). The black solid/dashed grey contours signal the $\pm$2,4,8..$\sigma$ levels. We show the synthesized beam in white in the left bottom of each panel. The rest-frame UV/optical position of the quasar is reported with a black cross. The continuum emission is not detected in J2228+0110 nor in J2053+0047.}
              \label{fig:contmap}%
    \end{figure*}

\begin{figure*}
   \centering
   \includegraphics[width=0.95\hsize]{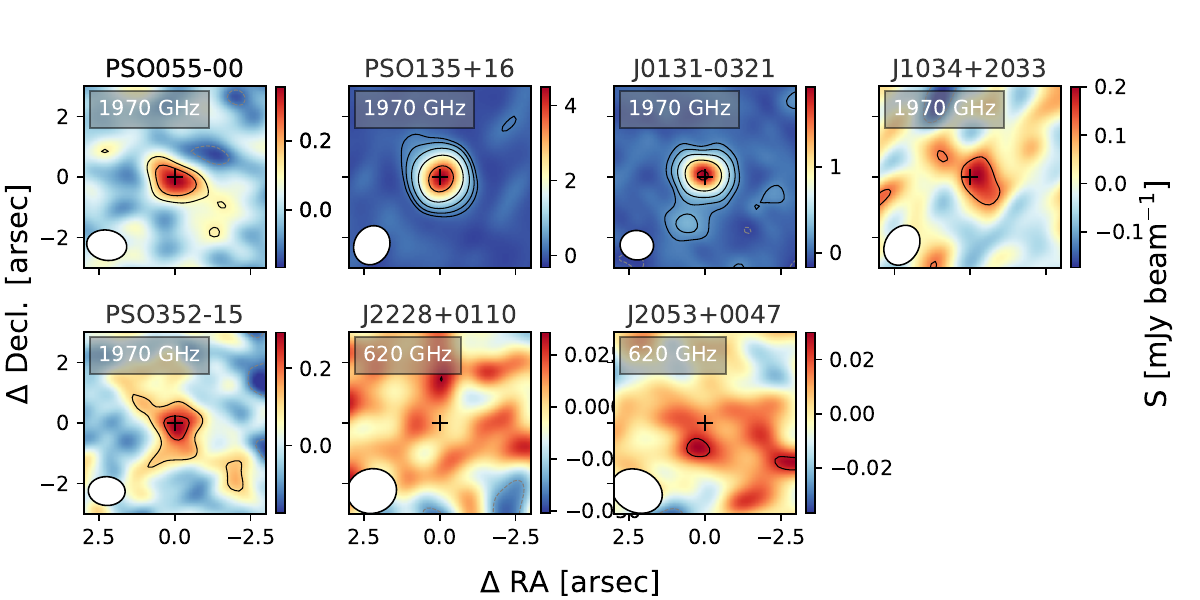}
   \caption{ALMA continuum emission maps at rest-frame 1970 GHz (Band 7) and 620 GHz (Band 3). Symbols as in Figure \ref{fig:contmap}}
              \label{fig:cont23map}%
    \end{figure*}

We report the continuum flux density and size values for observed Band 7 emission in Table \ref{tab:2dfit}, and limits on non detections in Band 3 in Table \ref{tab:co65}. The continuum maps are shown in Figures \ref{fig:contmap} and \ref{fig:cont23map}. For the remainder of the paper, to calculate the corresponding IR quantities (see Section \ref{sec:analysis}), we use the continuum emission flux obtained at 1900 GHz from the 2D Gaussian fit ($F_{\rm cont, 1900GHz}$), and reported in Table \ref{tab:2dfit}.
    
%
\begin{table*}
\caption{Results of the 2D Gaussian Fit of the continuum maps and the \cii\, continuum-subtracted emission line maps for the sources observed in Band 7.}             
\centering          
\label{tab:2dfit}
\begin{tabular}{l c c c c c c c}     
\hline\hline       
Name & Size$\rm _{\cii}$ & Size$_{\rm 1900 GHz}$ & Size$_{\rm 1970 GHz}$ & $F_{\rm \cii}$ & $F_{\rm \cii,corr}$ &  $F_{\rm cont, 1900 GHz}$ &  $F_{\rm cont, 1970 GHz}$ \\
     & [$^{\prime \prime}$] & [$^{\prime \prime}$] & [$^{\prime \prime}$] & [Jy km s$^{-1}$] &  [Jy km s$^{-1}$] & [mJy] & [mJy] \\
\hline                    
   J0131$-$0321 & 1.660 $\times$ 1.140 & 1.120 $\times$ 0.982 & 1.205 $\times$ 1.035 & 0.398$\pm$0.079 & 0.480$\pm$0.095 & 2.130$\pm$0.056 & 2.176$\pm$0.090 \\
   PSO055$-$00$^{\star}$  & 1.314 $\times$ 1.148 & 1.713 $\times$ 1.300 & 1.566 $\times$ 0.908 & 0.281$\pm$0.023 & 0.339$\pm$0.028 & 0.586$\pm$0.027 & 0.386$\pm$0.043 \\
   PSO135+16 & 1.352 $\times$ 1.261 & 1.352 $\times$ 1.245 & 1.295 $\times$ 1.179 & 2.504$\pm$0.092 & 3.017$\pm$0.111 & 4.42$\pm$0.16 & 4.510$\pm$0.15 \\
   J1034+2033 & 1.560 $\times$ 1.120 & 1.044 $\times$ 0.867 & 1.52 $\times$ 1.12 & 0.345$\pm$0.034 & 0.416$\pm$0.041 & 0.168$\pm$0.030 & 0.231$\pm$0.045 \\
   PSO352--15 & 1.614 $\times$ 1.184 & 1.280 $\times$ 0.897 & 1.91 $\times$ 1.30 & 0.695$\pm$0.093 & 0.837$\pm$0.112 & 0.300$\pm$0.031 & 0.581$\pm$0.098 \\
\hline
\end{tabular}
\tablefoot{
The corrected flux is obtained by considering a 1.20 factor (see Section \ref{sec:ciimeasure}). The (not deconvolved) size measurements provided here are consistent with the beam sizes.
           $^{\star}$ We note that PSO055-00's line emission is recovered at the edge of the bandwith. Therefore, caution should be taken when considering its \cii\,emission line value and all derived quantities.
           }
\end{table*}

\section{Analysis} \label{sec:analysis}
We report here gas and dust properties derived from the measurements of our RL quasars sample. We also compare them with those of RQ quasars' hosts at similar redshift taken from the literature.
\subsection{Literature Sample} \label{sec:analysis_sample}
In this work we compare the properties of RL quasars host galaxies with those derived from the literature.

For the RQ sample, we consider sources from the following studies, which targeted their dust continuum and \cii\,emission line: \citealt{walter2009}, \citealt{wang2013} \citeyear{wang2016}, \citealt{willott2013b}, \citeyear{willott2015}, \citeyear{willott2017}, \citealt{mazzucchelli2017b}, \citealt{trakhtenbrot2017}, \citealt{decarli2018}, \citealt{izumi2018}, \citeyear{izumi2019}, \citeyear{izumi2021}, \citealt{venemans2020}, \citealt{nguyen2020}, \citealt{yang2020}, \citealt{eilers2020a}, \citeyear{eilers2021}. The total number of RQ sources is 76.

We also consider six additional RL quasars' hosts galaxies properties from the literature. Four of these come from \cite{khusanova2022}: J0309+2717 ($z=6.115$), J1427+3312 ($z=6.115$), J1429+5447 ($z=6.1845$) and PSO172+18 (a potentially very young radio source at $z=6.823$, \citealt{momjian2021}, for which only upper limits in both continuum and \cii\,emission lines have been obtained).
The fifth source is J2318-3113 ($z=6.44$), which was re-classified as a RL quasar by \cite{ighina2021}. Using new 888 MHz radio observations acquired with the Rapid ASKAP Continuum Survey (RACS;  \citealt{mcconnell2020}, \citealt{hale2021}) they obtained a new radio loudness parameter of R$_{4400} \sim 70$ and observed a potential flux density variability of a factor of $\sim$2 in one year. In a recent study, \cite{ighina2024} presented new, multi-wavelength (radio, optical/NIR, X-ray) data, which reinforced the scenario of a young radio source with a black hole mass of $\sim8 \times 10^{8}$ M$_{\odot}$. Here, we adopt $R_{2500} = 152$ and a spectral index of $\alpha=0.54$ reported by \citet{ighina2021}. J2318-3113 has been previously observed with ALMA by \cite{decarli2018} and \cite{venemans2020}. We here include J2318-3113 in our RL quasar sample, utilizing the measurements provided by \cite{venemans2020}.
We finally include a very recently discovered blazar at $z=6.9964$, J0410-0139 \citep{banados2024a}. We consider here the continuum and \cii\,emission line fluxes from ALMA Band 6 observations obtained in August 2022 \citep{banados2024b}.
For all the sources from the literature, we re-derive IR and \cii\,luminosities, SFRs, and dust masses consistently with the objects presented in this work, and as outlined below in Section \ref{sec:analysis_lum}.  
\subsection{Redshift estimations}
We estimate the redshifts of the quasars in our sample from the peak of the \cii\, emission line, and we report these values in Table \ref{tab:1dcii}.

Several studies (e.g.\,\citealt{venemans2017c}, \citealt{decarli2018}, \citealt{meyer2019}, \citealt{schindler2020}) highlight a systematic, large blueshift of the rest-frame UV emission lines (e.g.\, \civ, \mgii) with respect to lines arising from the ISM in the host galaxy, considered the rest-frame of the system. Moreover the $\lya$ emission line at high-redshift is known to provide a poor estimation of the quasars' redshift, due to effects introduced by the absorption of the intergalactic medium (IGM; e.g.\, \citealt{banados2016}). Here, we find that the blueshift for J0131-0321, the only quasar whose redshift was known from the \mgii\,emission line, is very small, i.e.\,$\Delta v_{\rm \mgii - \cii} = -127$ km s$^{-1}$. This is consistent with the median difference of \,$\Delta v_{\rm \mgii - \cii}$ observed in the literature for RQ quasars ($\sim -367/ -390$ km s$^{-1}$ at $z\sim5/6$, \citealt{nguyen2020} and \citealt{schindler2020}). However, the blueshift relative to our \cii\, redshifts for the cases of literature redshifts based on observations of the $\lya$\,emission line or from template fitting are much larger, with a maximum value of $-$2060 km s$^{-1}$ (for PSO055-00, which is indeed recovered at the edge of the bandwidth). We measure a mean, median and standard deviation for \,$\Delta v_{\rm \lya / temp - \cii}$ of $-1269$, $-770$, and $2271$ km s$^{-1}$, respectively.

\subsection{Radio and (sub-)mm Spectral Energy Distribution} \label{sec:sed}
\begin{figure*}
   \centering
   \includegraphics[width=0.9\hsize]{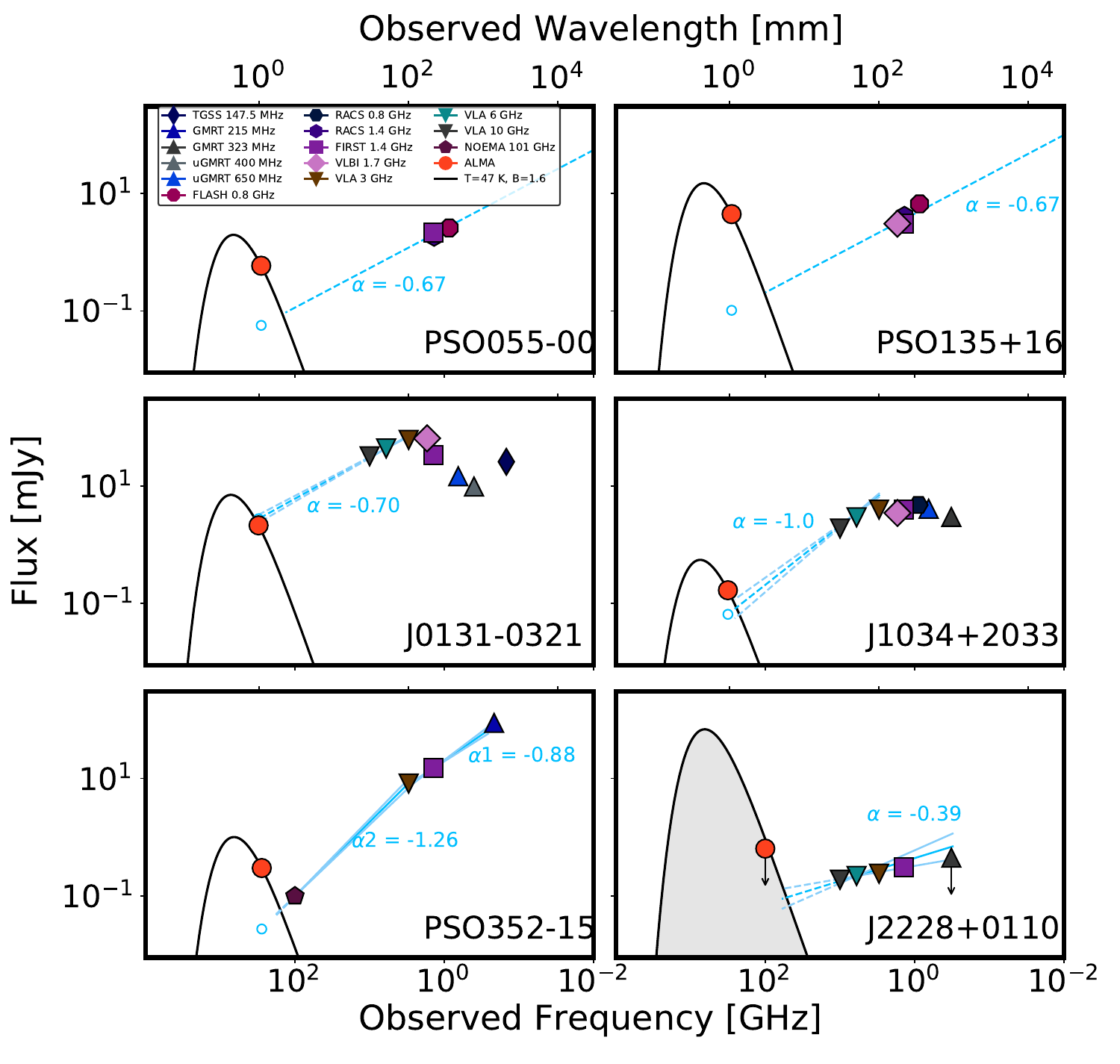}
   \caption{Radio and (sub-)mm Spectral Energy Distributions (SED) of RL quasars reported in this work. We show the continuum emission at 1mm from our ALMA Band 7 observations ({\it orange circles}), and the relative best fit modified black body dust emission ({\it black line}). In the case of J2228+0110, we could retrieve only upper-limits from our ALMA Band 3 data, and we show the corresponding limits on its dust emission with a {\it shaded grey area}. All modified black-body are calculated with T=47 K and $\beta=$1.6 (see Section \ref{sec:analysis_lum}). For each quasar, we report the radio observations obtained from a collection of literature data (see Section \ref{sec:sed} for references). For J0131-0321, J1034+2033, PSO352-15, J2228+0110 we show (broken) power-law radio emissions, with slopes ($\alpha$) derived from fits in the literature. For PSO055-00, PSO135+16, we assume a median high$-z$ slope value (see Section \ref{sec:sed}). The {\it solid light blue lines} are the measured power-law functions, while we report with {\it dashed lines} the extrapolation to lower/higher frequencies. The {\it light cyan} lines highlight the regions encompassed by the 1$\sigma$ uncertainties on the power-law slopes. We show with {\it light blue empty circles} the flux extrapolated from the synchtrotron power-law functions at the ALMA frequency targeted in this work.}
              \label{fig:sed}%
    \end{figure*}

We place our ALMA measurements of the continuum emission in the context of the spectral energy distribution (SED) of the RL quasars studied here. We utilize radio measurements from the literature (\citealt{zeimann2011}, \citealt{banados2018c}, \citealt{shao2020}, \citeyear{shao2022}, \citealt{rojas2021}, \citealt{krezinger2024}) obtained either via dedicated follow-up with VLA, VLBI, uGMRT and/or NOEMA, or from the FIRST and TGSS surveys. In addition, we obtained new measurements from RACS and the ASKAP First Large Absorption Survey in HI (ASKAP-FLASH; \citealt{allison2022}). In practice, we downloaded from CASDA\footnote{\url{https://data.csiro.au/domain/casda }} the available images of the quasars PSO055-00, PSO135+16 and J1034+2033, and we fitted a Gaussian profile within CASA to obtain the integrated fluxes at observed frequencies 0.89 GHz and 1.37 GHz (RACS) and at 0.86 GHz (FLASH). We note that the various reported observations were obtained in different times, within days to several years from each other, hence the radio intrinsic SED shape might be altered by variability effects. The resulting SEDs are shown in Figure \ref{fig:sed}.

In general, quasars' SEDs at radio frequencies are dominated by synchrotron emission, which can be modelled with a power law ($S_{\nu} \propto \nu^{\alpha}$).
In the case of PSO352-15, a broken power-law with $\alpha^{0.215}_{3} = -0.88 \pm 0.08$ (measured between observed frequencies 0.215 and 3 GHz) and $\alpha^{3}_{100} = -1.26 \pm 0.03$ (measured between observed frequencies 3 and 100 GHz) better reproduces the data; for a detailed discussion on this source, we direct the reader to \cite{rojas2021}. We highlight that extrapolating the sole synchrotron emission, one would obtain at the observed ALMA frequency a flux of $\sim$0.027 mJy (see Figure \ref{fig:sed}). Consequently, the estimated synchrotron contribution for this source is of $\sim$9\%, as discussed in \cite{rojas2021}.

J0131-0321 and J1034+2033 were extensively observed with the VLA and uGMRT by \cite{shao2020} and (\citeyear{shao2022}). A turn-over of the radio spectra was recovered, around an observed (rest-frame) frequency of $\sim$2 ($\sim$10) GHz in J0131-0321, and of $\sim$4 ($\sim$20) GHz in J1034+2033. Such radio spectral shapes are typically observed in GHz-peaked sources (GPS), or high-frequencies peakers (HFP). \cite{shao2022} shows that a model of free-free absorption by an external, in-homogeneous medium can best reproduce both cases. We report in Figure \ref{fig:sed} the corresponding, approximate power-law emission at frequencies higher than the turn-over, with the slopes indicated by \cite{shao2022}, which we can extrapolate to the frequencies of our dust emission model. We note that, in the case of J0131-0321, virtually most of the emission we recover from our ALMA continuum observations at observed 300 GHz could be due to synchrotron emission. For J1034+2033, the recovered intensity measured with ALMA (0.168 mJy; see Table \ref{tab:2dfit}) is higher than what is expected by the sole synchrotron emission extrapolation ($\sim$0.065 mJy; see Figure \ref{fig:sed}). In both cases, a mix of dust+synchrotron radiation could be responsible for the continuum emission observed with ALMA. In particular, extrapolating the radio power-law, we can obtain a first order estimate of synchrotron contribution at $\sim$300 GHz of $\sim$40\% and $\sim$100\% for J1034+2033 and J0131-0321, respectively.

J2228+0110 is the weakest radio emitter in this sample. It was observed with VLA and uGMRT by \cite{shao2020}, and (\citeyear{shao2022}): in contrast to J0131-0321 and J1034+2033, no clear radio turn-over was observed, although the upper limits recovered with the uGMRT could provide a hint of a change in the spectrum. Not considering this upper limit, the radio emission could be modelled, between observed frequencies 323 MHz and 1 GHz, with a relatively flat power law, with $\alpha=-0.39 \pm 0.17$. This object was also not detected in our ALMA Band 3 observations, hence we only have upper limits for its continuum emission, which we show with gray shades in Figure \ref{fig:sed}.

PSO055-00 and PSO135+16 were observed only within 0.8-1.7 GHz, hence a model of their radio emission can not be meaningfully fitted; we show in Figure \ref{fig:sed} an indicative power law emission with slope $\alpha=-0.67$, used in \cite{banados2021} to derive their radio-loudnesss parameters, and calculated as the median slope of the radio emission in $z>5$ quasars. In both cases, this slope is broadly consistent with the radio measurements, and we do not observe strong variability. Our ALMA measurements (0.586 mJy and 4.42 mJy for PSO055-00 and PSO135+16, respectively; see Table \ref{tab:2dfit}) are approximately one order of magnitude higher than the flux density obtained when extrapolating the synchrotron emission ($\sim$0.057 mJy and $\sim$0.102 mJy for PSO055-00 and PSO135+16, respectively; see Figure \ref{fig:sed}). In this case, the estimated synchrotron contribution at $\sim$300 GHz are $\sim$9\% and $\sim$2\% for PSO055-00 and PSO135+16, respectively. 

All these lines of evidence show the variety of the SEDs of high$-z$ RL quasars, and highlights how observations at rest frame $\sim$1mm, in case of RL quasars, could be the result of thermal and non-thermal emission. Therefore, extreme caution should be used when deriving SFR, dust/gas masses etc.\, from continuum measurements at these wavelengths. In order to still provide valuable estimates, discussions and comparisons with the literature, while considering the above described limitations with the currently available datasets, we derive such quantities in two ways. We firstly consider the total ALMA recovered continuum data as purely produced by dust, which will provide upper limits to IR luminosities, SFR and dust/gas masses. Secondly, we derive these measurements subtracting the synchrotron values as extrapolated at $\sim$1mm from the radio power-law functions (as described above) to the total observed ALMA continuum flux. Considering that the radio power-law functions are expected to bend downward towards higher frequencies, due to the interaction between high-energy particles in the jets and the surrounding medium (e.g.\,\citealt{jaffe1973}), decreasing the synchrotron contribution, the IR-related quantities derived in this way can be considered as lower limits. The intrinsic SFR, IR luminosities and gas/dust masses should then be found within the two reported values. Given that virtually $\sim$100\% of the $\sim$300 GHz flux density of J0131-0321 could be due to synchrotron, we do not derive a second measure for this source, but consider the firstly derived IR values as pure upper limits. Also, we do not derive a second set of values for J2228+110, as only upper limits are recovered from our ALMA data.
Further observations at intermediate frequencies are needed to properly model these RL quasars SED, and to more accurately and quantitatively retrieve all their components.

\subsection{IR and \cii\, luminosities, Star Formation Rates and Dust Masses} \label{sec:analysis_lum}
We can estimate the \cii\, line luminosity from the measured velocity-integrated line flux, following \cite{carilli2013} :
\begin{equation} \label{eq:lcii}
    \dfrac{L_{\rm [CII]}}{L_{\odot}} = 1.04 \times 10^{-3} \dfrac{F_{\rm [CII]}}{\rm Jy\, km\, s^{-1}} \dfrac{\nu_{\rm obs}}{\rm GHz} \left(\dfrac{D_{\rm L}}{\rm Mpc}\right)^{2}
\end{equation}
with $F_{\rm [CII]}$ flux of the \cii\, emission line, $\nu_{\rm obs}$ the observed frequency of the line, and $D_{\rm L}$ the luminosity distance.

Meanwhile, the infrared emission of high$-z$ quasars can be modelled with a modified black body (e.g., \citealt{decarli2018}, \citealt{venemans2018}). Hence, in the optically thin scenario, the observed continuum flux density can be derived as follows (e.g.\,\citealt{novak2019}): 
\begin{equation}
    F_{\rm obs} = \dfrac{f_{\rm CMB} [1+z]}{D_{\rm L}^{2}} \kappa_{\nu,{\rm rest}}(\beta) M_{\rm dust} B_{\nu,{\rm rest}}(T_{\rm dust}, z)
\end{equation}
where $M_{\rm dust}$ is the dust mass, $T_{\rm dust}$ is the dust temperature and $B_{\nu,{\rm rest}}(T_{\rm dust}, z)$ is the Planck function:
\begin{equation}
    B_{\nu,{\rm rest}}(T_{\rm dust},z) = \dfrac{2h}{c^{2}} \nu_{\rm rest} \left[ \exp \left( {\dfrac{h_{\rm \nu}}{k T_{\rm dust}}}\right) -1 \right]
\end{equation}
The factor  $\kappa_{\nu,{\rm rest}}(\beta)$ is the dust mass opacity coefficient, which can be written as (e.g.\,\citealt{dunne2000}):
\begin{equation}
    \kappa_{\nu,{\rm rest}}(\beta) = 0.077 \left( \dfrac{\nu_{\rm rest}}{\rm 352 GHz} \right)^{\beta} {\rm m^{2} kg^{-1}}
\end{equation}
with $\beta$ the dust emissivity power law spectral index. At high redshift ($z\gtrsim5$) the Cosmic Microwave Background (CMB) contrast starts to become significant. We can take into account this correction via the factor $f_{\rm CMB}$ (e.g.\,\citealt{dacunha2013}):
\begin{equation}
    f_{\rm CMB} = 1 - \dfrac{B_{\nu, {\rm rest}} (T_{{\rm CMB},z})}{B_{\nu, {\rm rest}} (T_{{\rm dust},z})}
\end{equation}
Here, we consider values for dust temperature and $\beta$ typically used in the literature, i.e.\, $T_{\rm dust} = 47 K$ and $\beta = 1.6$ (e.g.\,\citealt{beelen2006}, \citealt{decarli2018}, \citealt{rojas2021}). 
We scale the above function to the observed continuum flux densities at $\nu_{\rm rest}=1900$ GHz of our sources (as derived from the 2D Gaussian fit), as well as considering the synchrotron corrected ones (see Section \ref{sec:sed}). Hence, we derive dust masses for our quasars' hosts of $<$3$\times$ 10$^{8} M_{\odot}$.

We calculate the far-infrared luminosity ($L_{\rm FIR}$) by integrating the modified black body between rest-frame wavelength 42.5 and 122.5 $\mu$m (e.g.\, \citealt{helou1988}). We also estimate the total infrared luminosity ($L_{\rm TIR}$) by integrating the SED between 8 and 1000 $\mu$m (e.g.\, \citealt{kennicutt2012}), which is equivalent to $L_{\rm FIR} = 0.75 L_{\rm TIR}$ for the model considered here.

We also estimate the equivalent width of the \cii\, emission line ($\rm EW_{[CII]}$):
\begin{equation}
    \dfrac{\rm EW_{[CII]}}{\rm km\, s^{-1}} = 1000 \times \dfrac{F_{\rm [CII]}\, [{\rm Jy\, km\, s^{-1}}]}{F_{\rm cont}\, {\rm [mJy]}} 
\end{equation}

We can compute the star formation rates in the quasars' host galaxies by relying both on the \cii\, emission line ($\rm SFR_{[CII]}$) and the infrared luminosities ($\rm SFR_{IR} $). 
Several studies establish a relation between $L_{\rm \cii}$ and star formation rates, e.g.\,\cite{delooze2011}, \cite{sargsyan2012}, \cite{herreracamus2015}. Here, we obtain \cii-derived SFR using:
\begin{equation}
    {\rm SFR_{[CII]}} = 3 \times 10^{-9} \left( \dfrac{L_{\rm [CII]}}{L_{\odot}} \right)^{1.18}
\end{equation}
from \cite{delooze2014}, whose relation was calibrated on $z>0.5$ galaxies with a scatter of $\sim$0.4dex. 
For the dust-based star formation rates we use the equation from \cite{kennicutt2012}:
\begin{equation}
    \dfrac{\rm SFR_{IR}}{M_{\odot}\, {\rm yr^{-1}}} = 1.49 \times 10^{-10} \dfrac{L_{\rm TIR}}{L_{\odot}}
\end{equation}

We report the $L_{\rm [CII]}$, $\rm EW_{[CII]}$ and $\rm SFR_{\cii}$ values obtained for the RL quasars in this work in Table \ref{table:analysis_cii}.
The $L_{\rm FIR}$, $M_{\rm dust}$ and $\rm SFR_{IR}$ for the RL quasars newly reported here, and calculated considering the two flux density sets as described in Section \ref{sec:sed}, are listed in Table \ref{table:analysis_cont}.

\begin{figure}
   \centering
   \includegraphics[width=\hsize]{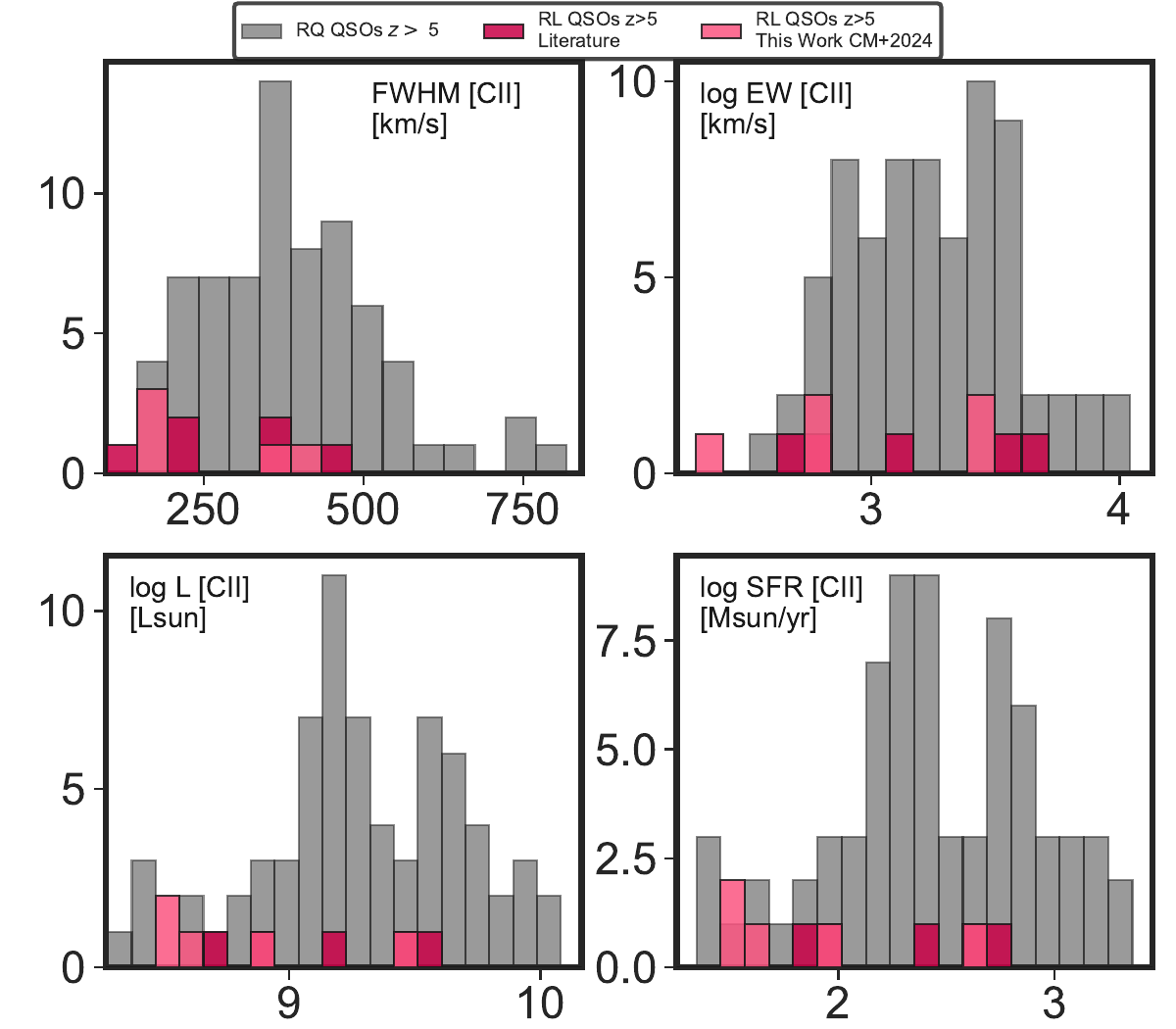}
   \caption{ Distribution of \cii\, emission line and properties for RL (\textit{pink}), and RQ (\textit{grey}) quasars. }   
              \label{fig:distrprop}%
    \end{figure}

In Figure \ref{fig:distrprop}, we show the distribution of $L_{\rm [CII]}$, $\rm EW_{[CII]}$, $\rm FWHM_{[CII]}$ and $\rm SFR_{\cii}$, and their comparison with the literature sample of RQ and RL quasars. We also report the mean, median and standard deviation of the two samples in Table \ref{table:mean}.
Despite the limited data for RL quasars with respect to the RQ ones, we can already note that the distribution of \cii\,emission line EW and FWHM are similar between the two samples. We performed a Kolmogorov-Smirnov (KS) test in order to assess if these properties are taken from the same underlying distribution, and report the corresponding p-values in Table \ref{table:mean}. For p-values $<0.05$, we can reject the null hypothesis (i.e.\,reject that the underlying distribution is the same). In our case, we calculate a p-value of 0.24 and 0.14 for EW and FWHM distributions, respectively, indicating that there are no systematic differences between RL and RQ quasars. Finally, we can observe potential differences in the distribution of \cii\,emission line luminosities, where RL quasars seem to be systematically fainter than the RQ ones. We obtain a p-value of 0.04, moderately suggesting that there are differences in the two samples. More observations of \cii\,emission in a larger sample of RL quasars at high-redshift will allow us to explore this hypothesis further.

\begin{figure}
   \centering
   \includegraphics[width=\hsize]{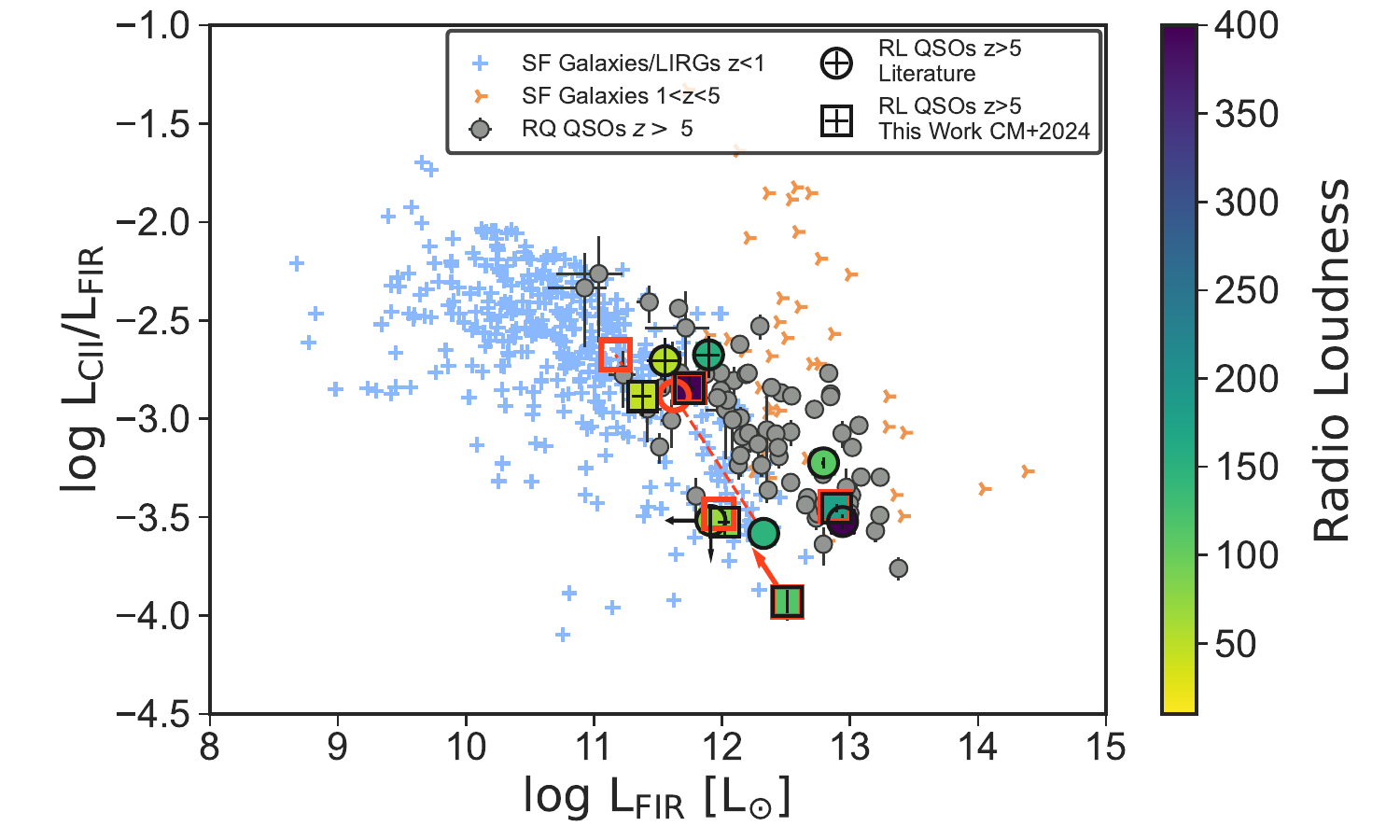}
   \caption{\cii-to-FIR luminosity ratio as a function of FIR luminosity. Observations of star-forming galaxies and LIRGs at $z<1$ are shown with \textit{light blue crosses}, while we report star-forming galaxies at $1<z<5$ with \textit{orange arrows}. RQ quasars from the literature are depicted with \textit{grey points}. All references for these data are reported in Section \ref{sec:analysis_sample} and \ref{sec:analysis_lum}. We show in \textit{circles} RL quasars observations from \cite{khusanova2022}, \cite{venemans2020} and \cite{banados2024b} and in \textit{squares} the values derived in this work, considering that the total measured ALMA continuum flux is due only to dust, all color-coded for radio loudness parameter. We note that RL quasars occupy a similar parameter space of that of RQ quasars' hosts, with the exception of the sources J0131-0321 and J0410-0139 (with $L_{\rm \cii} / L_{\rm FIR} \sim 10^{-3.5-4}$). When considering FIR luminosities accounting for the synchrotron contribution-derived flux ({\it red squares} for sources from this work, {\it red circle} from \citealt{banados2024b}), even the above mentioned outliers move closer to the RQ quasars' hosts location. }   
              \label{fig:lfir_lciifir}
    \end{figure}

An important effect observed in a variety of galaxies and AGN at high and low redshift is the so-called ``\cii\,deficit'', i.e.\,a decrease of the $L_{\rm \cii}/L_{\rm FIR}$ ratio as a function of infrared luminosity. A unique understanding of the origin of this effect is still lacking, as there are several factors that could contribute to the suppression of \cii\,luminosity. Some scenarios considered in the literature are: the ionization of \cp due to X-ray radiation from an AGN, that also boost FIR luminosities (e.g.\,\citealt{langer2015}); self-absorption of \cii\,in high density environments, which becomes higher than the critical density for collisional interaction (e.g.\citealt{diazsantos2017}, \citealt{sutter2021}); or thermal saturation of \cp due to strong far-UV radiation (e.g.\,\citealt{rybak2019}). In Figure \ref{fig:lfir_lciifir} we show $L_{\rm \cii}/L_{\rm FIR}$ vs $L_{\rm FIR}$ for different samples: star-forming galaxies and Luminous Infrared Galaxies (LIRGs) at $z<1$ (\citealt{farrah2013}, \citealt{sargsyan2014}, \citealt{malhotra2001} and \citealt{diazsantos2013}), star forming galaxies at $1<z<5$ (\citealt{stacey2010}, \citealt{brisbin2015}, \citealt{gullberg2015}), and RQ and RL quasars at $z>5$. In the low-redshift universe, there is a clear decrement of \cii\,luminosity with increasing FIR luminosity, while for high$-z$ star forming galaxies the scatter is larger. RQ quasars hosts at $z>5$ follow a trend similar to low$-z$ ULIRGs/star forming galaxies, with $L_{\rm \cii}/L_{\rm FIR}$ between $\sim 10^{-2} - 10^{-3}$.  The RL quasars' hosts mainly follow the distribution of RQ ones, with no strong dependency on the radio loudness parameter. We note that, if we consider the entire $\sim$1mm flux density as due to dust, the quasars J0131-0321 and J0410-139 show the largest difference to the parameter space occupied by high$-z$ quasars, with a $L_{\rm \cii}/L_{\rm FIR}$ ratios as low as 2.5-1.2$\times$10$^{-4}$. However, once considering the synchrotron contribution, these sources move closer to the locus of RQ quasars' hosts.
\subsection{IR and CO(6-5) luminosities limits} \label{sec:co65analysis}
We estimate limits on the \co\, emission line luminosities ($L_{\rm \co}$) using Equation \ref{eq:lcii}. One can also express the line luminosity via the quantity $L'$ in units of [K km s$^{-1}$ pc$^{2}$] following \cite{carilli2013}:
\begin{equation}
    L'_{\rm \co} = 3.25 \times 10^{7} \times  \dfrac{F_{\rm \co}}{\rm Jy\, km\, s^{-1}} \dfrac{1}{(1+z)^{3}} \left(\dfrac{\nu_{\rm obs}}{\rm GHz}\right)^{-2} \left(\dfrac{D_{\rm L}}{\rm Mpc}\right)^{2}
\end{equation}
Additionally, we can estimate the limits on the far and total-infrared luminosities as done for the objects observed in Band 7 (see Section \ref{sec:ciimeasure}). All the limits on \co\, and infrared luminosities from Band 3 observations are reported in Table \ref{tab:co65}.

\begin{figure}
   \centering
   \includegraphics[width=0.75\hsize]{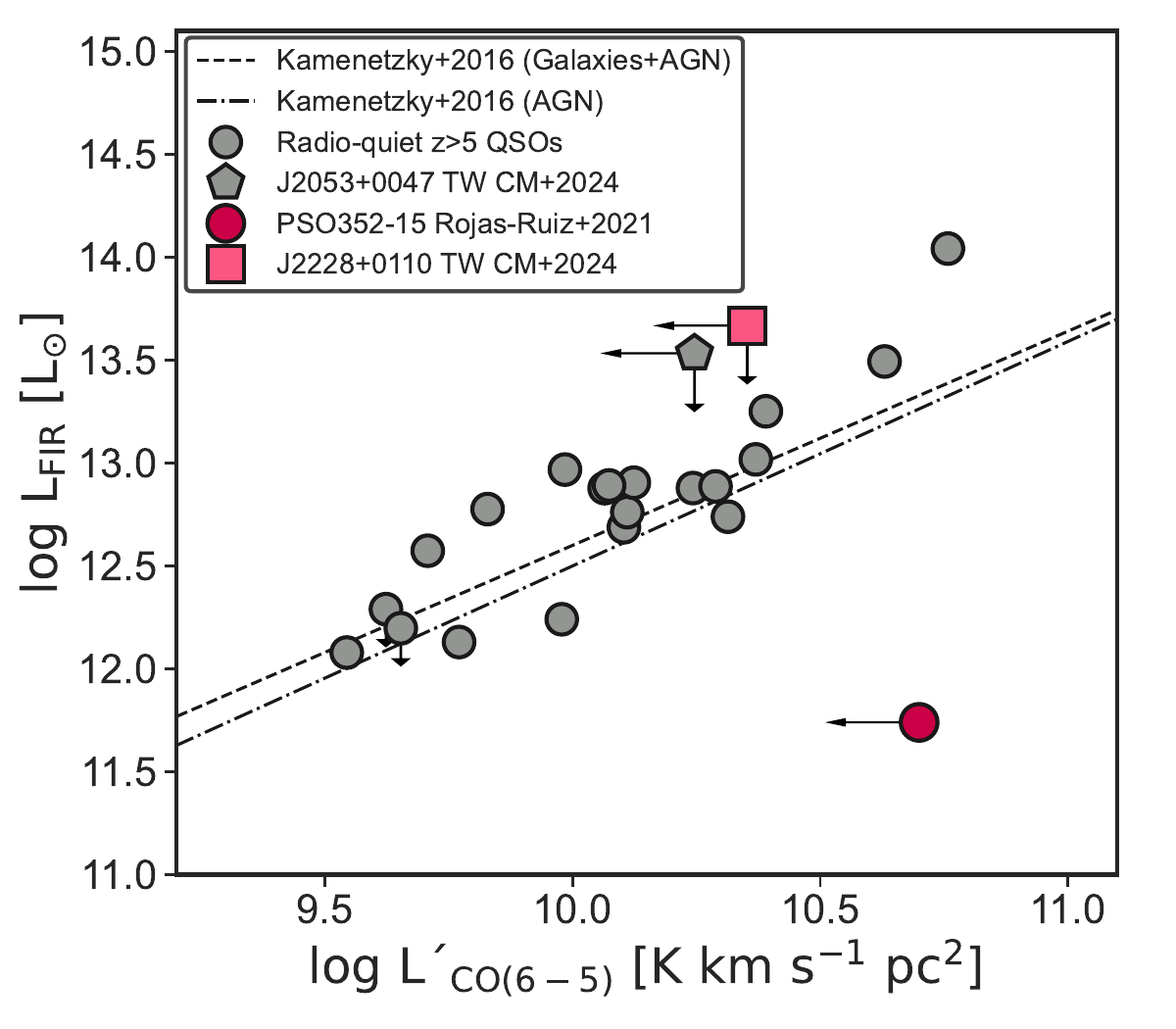}
   \caption{Far-infrared (42-122.5 $\mu$m) luminosity vs \co\, luminosity for the two sources observed in Band 3 in our sample (J2053+0047, RQ, {\it grey penthagon}; J2228+0110, RL, {\it pink square}), compared to a sample of RQ $z>5$ quasars in the literature ({\it grey circles}, see text for references). Measurements and limits for the RL quasar PSO352-15 are also shown (\citealt{rojas2021}; {\it dark pink circle}). The relations measured for galaxies+AGN, and only AGN, at lower$-z$, in \cite{kamenetzky2016} are also reported ({\it dashed} and {\it dot-dashed black lines}). The sources in this work are consistent with values in the literature, particularly given the relatively loose limits on both the infrared and \co\, emission line luminosities.}
              \label{fig:lco65_lfir}
    \end{figure}

We compare the upper limits found here with \co\,emission line observations in a sample of RQ quasars at $z>5$ from the literature (\citealt{wang2010}, \citeyear{wang2013}, \citealt{feruglio2018}, \citeyear{feruglio2023}, \citealt{wangf2019}, \citealt{yangj2019}, \citealt{decarli2022}, \citealt{kaasinen2024}). We also consider the source PSO352-15, part of the sample here. This source was observed with NOEMA at 101 GHz by \cite{rojas2021}. They estimated a 3$\sigma$ limit on the \co\,emission line of $L'_{\rm \co} < 5 \times 10^{10}$ K km s$^{-1}$ pc$^{2}$ (see their Section 3.2 for further details). Conversely, we consider the corresponding FIR luminosity derived in this work (see Table \ref{table:analysis_cont}), which is consistent with the value reported by \cite{rojas2021}.
We report $L_{\rm FIR}$, re-calculated for the RQ sources in the literature with the method reported in Section \ref{sec:analysis_lum} for consistency, as a function of \co\,emission line luminosities in Figure \ref{fig:lco65_lfir}. We also report the expected relation between these two quantities calculated in \cite{kamenetzky2016}. The limits on these quantities for the two objects newly presented here do not show a critical difference between them and the RQ quasars sample. Given the somewhat loose constraints on the \co\, luminosity of PSO352-15, this source is also consistent with the RQ quasars measurements. Deeper observations on a larger sample of RL quasars are needed to place more meaningful constraints. 
\subsection{Mass Budget} \label{sec:analysis_massdyn}
A key physical parameter that one can derive from sub-mm observations of high$-z$ quasars' hosts is the dynamical mass of the system. This, in combination with measurements of the central SMBH mass, allows for studies of the SMBH/galaxy co-evolution at very high redshifts. Based on such comparisons, several works have shown that SMBHs in the EOR seem to be overmassive with respect to systems in the local universe (e.g.\, \citealt{farina2022}; see \citealt{pacucci2024} for a recent compilation, but also \citealt{izumi2021}).

In this work, the FIR continuum and \cii\,emission of the radio-loud quasars' host galaxies are unresolved (see Figure \ref{fig:1dciimap}, \ref{fig:ciimapmom0} and Table \ref{tab:2dfit}).
Hence, the quality of our data prevents us from performing a full modeling of the kinematics of our sources, as done in studies of $z>5$ quasars' hosts with higher spatial resolution and SNR observations (e.g.\, \citealt{pensabene2020}, \citealt{neeleman2021}). Therefore, we estimate the dynamical masses of the host galaxies presented here in a simplified way, following \cite{decarli2018}. Briefly, under the assumption that the system is dominated by dispersion, one can express $M_{\rm dyn}$ with the following: 
\begin{equation}
    M_{\rm dyn} = \frac{3}{2} \dfrac{ {\rm R_{\cii}} \sigma_{\rm \cii}^{2} }{\rm G} = \frac{\rm R_{\cii}}{G} \left( \dfrac{\rm 0.75 FWHM_{\cii}}{sin(i)} \right)^{2}
\end{equation}
where $\rm R_{\cii}$ is the size of the \cii\,emission line, considered as the major axis of the 2D Gaussian fit of the \cii\,map, $G$ is the gravitational constant and $\sigma_{\rm \cii}$ is derived from the Gaussian fit of the line, which is equivalent to 0.75$\times$FWHM$_{\rm \cii}$. The second equation stands valid assuming that the width of the line is rotation-dominated, considering a flat disk structure with an inclination angle $i$. We utilize an angle $i=46^{\circ}$, i.e.\, the median inclination of a $z>6$ quasar hosts' sample derived by \cite{wangf2024}.
We obtain dynamical masses between $2-20 \times 10^{10} M_{\odot}$ (see Table \ref{table:analysis_cii}). This is broadly consistent with values similarly obtained for RQ quasars in the literature (e.g.\, \citealt{decarli2018}, \citealt{wangf2024}).

We can also infer molecular masses ($M_{\rm H2}$) relying on the \cii\,luminosity, as calibrated from a sample of $z\sim2$ main sequence and starburst galaxies by \cite{zanella2018}:
\begin{equation}
    \dfrac{M_{\rm H2, \cii}}{M_{\odot}} = \dfrac{\alpha_{\rm \cii}}{M_{\odot}\, L_{\odot}^{-1}} \dfrac{L_{\rm \cii}}{L_{\odot}}
\end{equation}
with $\alpha_{\rm \cii} = 30 M_{\odot} L_{\odot}^{-1}$. We report the values of $M_{\rm H2,\cii}$ for the sources presented here in Table \ref{table:analysis_cii}. 
RL sources do not show significant differences with respect to RQ ones (e.g.\,\citealt{decarli2022}). 
\begin{table*}
\label{tab:sample}
\caption{ \cii\, luminosities, emission line equivalent widths, \cii\,-based star formation rates, dynamical and \cii\,-based molecular gas masses values for the RL quasars in our sample. }             
\label{table:analysis_cii}      
\centering          
\begin{tabular}{l c c c c c}     
\hline\hline       
Name & $L_{\rm [CII]}$ & EW$\rm _{[CII]}$ & $\rm SFR_{[CII]}$ & $M_{\rm dyn}$ & $M_{\rm H2,\cii}$ \\
     & [$10^{8}\, L_{\odot}$] & [km\, s$^{-1}$] & [$M_{\odot}\, {\rm yr}^{-1}$] & [$10^{10}\, M_{\odot}$]  & [$10^{9}\, M_{\odot}$] \\
\hline                    
   J0131-0321 & 3.8$\pm$0.8 & 225$\pm$45 & 40$\pm$9 & 4.8$\pm$4.0 & 11$\pm$2.0 \\ 
   PSO055-00$^{\star}$  & 3.1$\pm$0.3 & 578$\pm$78 & 32$\pm$3 & 2.8$\pm$2.2 & 9.3$\pm$0.8\\ 
   PSO135+16 & 27.7$\pm$1.0 & 683$\pm$35 & 416$\pm$18 & 15.0$\pm$1.8 & 83$\pm$3.0\\ 
   J1034+2033 & 3.1$\pm$0.3 & 2476$\pm$505 & 32$\pm$4 & 5.8$\pm$4.9 & 9.3$\pm$0.9 \\ 
   PSO352--15 & 7.9$\pm$1.1 & 2790$\pm$472 & 95$\pm$15 & 23.3$\pm$10.7 & 24$\pm$3.0 \\ 
\hline
\end{tabular}
\tablefoot{
           \textbf{$^{\star}$} We note that PSO055-00's line emission is recovered at the edge of the bandwith. Therefore, caution should be taken when considering its \cii\,emission line value and all derived quantities.
           }
\end{table*}
%
\begin{table*}
\caption{ Mean, median and standard deviation of the \cii\,emission line properties shown in Figure \ref{fig:distrprop}, for the RQ and RL high$-z$ quasars samples.}             
\label{table:mean}      
\centering          
\begin{tabular}{l c c c c c c c}     
\hline\hline       
 & \multicolumn{3}{c}{RQ quasars} & \multicolumn{3}{c}{RL quasars$^{\star}$} & KS-test\\
 & mean & median & st.dev & mean & median & st.dev & p-value \\
\hline                    
FWHM$_{\rm [CII]}$ [km s$^{-1}$]       &  387 & 379  & 133  &  272 & 290 & 110 & 0.14 \\
EW$_{\rm [CII]}$ [km s$^{-1}$]         & 2362 & 1691 & 2013 & 1848 & 1180 & 1441 & 0.24 \\
$L_{\rm [CII]}$ [$10^{9}\, L_{\odot}$] & 2.9  & 1.8  & 2.6  & 1.5 & 0.8 & 1.2 & 0.04 \\
$\rm SFR_{[CII]}$ [M$_{\odot}\,{\rm yr}^{-1}$] & 466 & 258 & 498 & 211 & 95 & 194 & 0.04 \\
\hline
\end{tabular}
\tablefoot{
           We also report the p-values obtained with a Kolmogorov-Smirnov test, to assess if the observed quantities for RQ and RL quasars were obtained from the same underlying distribution.
           \textbf{$^{\star}$} We only consider sources with \cii\,emission lines detection, i.e. we exclude PSO172+18 from this comparison.
           }
\end{table*}
\begin{table*}
\caption{ FIR Luminosities, FIR-based star formation rates and dust masses for the RL quasars in our sample, calculated considering the total measured ALMA flux density ({\it tot}) and by correcting for the extrapolated synchrotron flux ({\it syn}). }       
\label{table:analysis_cont}      
\centering          
\begin{tabular}{l c c c c c c }     
\hline\hline       
Name & $L_{\rm FIR,tot}$ & $L_{\rm FIR,syn}$ & $\rm SFR_{IR,tot}$ & $\rm SFR_{IR,syn}$ & $M_{\rm dust,tot}$ & $M_{\rm dust,syn}$ \\
     & [$10^{11}\, L_{\odot}$] & [$10^{11}\, L_{\odot}$] & [$M_{\odot}\, {\rm yr}^{-1}$] & [$M_{\odot}\, {\rm yr}^{-1}$]  & [$10^{7}\, M_{\odot}$] & [$10^{7}\, M_{\odot}$]  \\
\hline                    
   J0131-0321\textbf{$^{\dagger}$} & 32.4$\pm$0.85 & - & 400$\pm$10 & - & 19.9$\pm$0.3 & - \\ 
   PSO055-00  & 10.5$\pm$1.1 & 9.4$\pm$1.1 & 129$\pm$14 & 116$\pm$14 & 2.8$\pm$2.2 & 3.8$\pm$0.4 \\ 
   PSO135+16  & 78.8$\pm$2.9 & 77$\pm$2.8 & 973$\pm$35 & 951$\pm$35 & 31.5$\pm$1.1 & 30.7$\pm$1.1 \\ 
   J1034+2033 & 2.4$\pm$0.4 & 1.47$\pm$0.4 & 30$\pm$5 & 18$\pm$5 & 0.9$\pm$0.2 & 0.6$\pm$0.2 \\ 
   PSO352--15 & 5.5$\pm$0.6 & 5.5$\pm$0.8 & 68$\pm$7 & 68$\pm$7 & 2.2$\pm$0.2 & 2.2$\pm$0.2 \\ 
\hline
\end{tabular}
\tablefoot{
           \textbf{$^{\dagger}$} The synchrotron contribution could amount to almost all the flux density at $\sim$300 GHz for this source. Hence, we report in this case only the IR values obtained from considering the total measured ALMA flux density as due to the dust, and use them as upper limits.
           }
\end{table*}
\subsection{Impact of AGN Luminosity on \cii\,and FIR Emission}
Unveiling the relative role of AGN vs star formation in heating gas and dust in the ISM is crucial to understand quasar/galaxy (co)evolution, especially at high redshift.

This can be done by comparing  observations of different far-infrared emission lines (e.g.\,[N\,II]\,205$\mu$m, [O\,I]\,146$\mu$m, [O\,III]\,88$\mu$m, [C\,I]\,369$\mu$m) with output of photoionization codes (e.g.\,CLOUDY; \citealt{ferland2017}).
Some works in the literature provide these studies for few $z\gtrsim6$ quasars, showing that their host galaxies are mainly heated by star formation, with apparently no evident contribution from the AGN (e.g.\,\citealt{venemans2017b}, \citealt{novak2019}, \citealt{pensabene2021}, \citealt{meyer2022a}).

However, such investigations are currently limited to few objects, due to the time consuming nature of the observations of different, faint lines in the (sub-)mm. Hence, to gauge the AGN contribution to the host galaxy in a larger sample of quasars, one can resort to more straightforward methods, e.g.\,comparing the absolute UV magnitude at rest frame wavelength 1450 $\AA$, linked with the AGN activity, to the FIR and \cii\,luminosities, arising instead from the host galaxy. \cite{decarli2018} and \cite{venemans2018} provided this comparison for $\sim30$ UV-bright quasars observed with ALMA, remarking that no strong correlation between UV and FIR was recovered, especially when focusing on the more complete sample at high UV luminosities. \cite{venemans2020} confirmed this lack of correlation with ALMA observations at higher ($\sim$kpc) spatial resolution, concluding that not even the most central FIR emission seemed to be affected by the presence of an AGN.

\begin{figure}
   \centering
   \includegraphics[width=0.95\hsize]{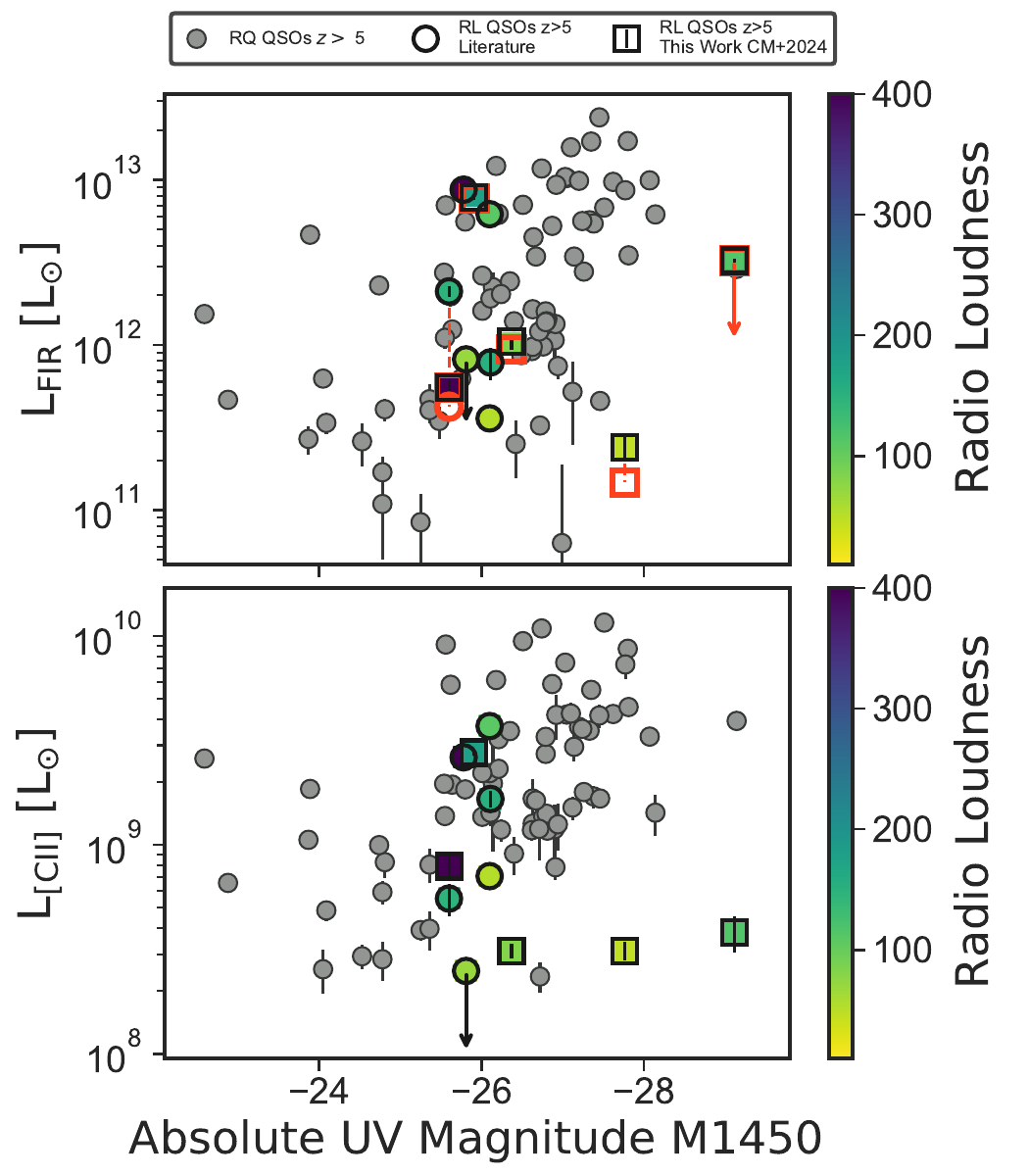}
   \caption{Far-infrared (\textit{top}) and \cii\, (\textit{bottom}) luminosity vs absolute UV magnitude at rest-frame wavelength 1450 \AA. RQ quasars from the literature (see Section \ref{sec:analysis_sample} for references) and RL quasars are shown with the same symbols as in Figure \ref{fig:lfir_lciifir}.}
   \label{fig:M1450IR}%
    \end{figure}

Here, we show $L_{\rm \cii}$ and $L_{\rm FIR}$ as a function of $M_{1450}$ for the RQ quasars sample from the literature and for the RL quasars (Figure \ref{fig:M1450IR}). We note that the majority of the RL quasars fall in the same parameter space as the RQ ones, i.e.\, in both samples, considering a given UV luminosity, we observe a large range of $L_{\rm \cii}$ and $L_{\rm FIR}$ values, with a scatter larger than an order of magnitude. No strong correlation with radio-loudness can be recovered. However, we note that few objects (J0131-0321, J1034+2033, PSO055-00 and PSO172+18) are outliers in the $L_{\rm \cii}$ vs $M_{1450}$ plane, i.e.\,they show a much lower \cii\,luminosity than that observed in RQ quasars with a comparable UV luminosity.


\subsection{Companions}  \label{sec:analysis_comp}
RL quasars and radio galaxies are found to be surrounded by overdensities of galaxies (e.g.\,\citealt{wylezalek2013}, \citealt{gilli2019}). Different studies using rest-frame UV observations recovered a rich environment around the $z\sim5.8$ RL quasar J0836+0054 (e.g.\,\citealt{zheng2006}, \citealt{ajiki2006}, \citealt{bosman2020}, \citealt{overzier2022}). At the same time, ALMA observations of RQ quasars found several, line-emitters in their surroundings, from systems in advanced mergers to galaxies at tens of kpc distances (e.g.\,\citealt{decarli2017}, \citealt{trakhtenbrot2017}, \citealt{nguyen2020}, \citealt{neeleman2021}). Nevertheless, we note that studies looking for continuum dust emitters around quasars were not successful in discovering clear overdensities (e.g.\,\citealt{champagne2018}, \citealt{meyer2022b}). 

In this work, we search for line emitters in the fields of the RL quasars observed in Band 7, using the continuum emission subtracted cubes obtained as reported in Section \ref{sec:ciimeasure}. We utilize the code \textit{findclumps}\footnote{With the python implementation provided in the interferopy package: \url{https://interferopy.readthedocs.io/en/latest/index.html}, \cite{boogaard2021}.} \citep{walter2016}. In brief, $\textit{findclumps}$ performs a floating average of channels over different spectral windows. For each window, it searches for positive and negative emission peaks at different SNR, where the noise is calculated as the rms of the collapsed map. In this work, we use spectral windows of width between 3 and 19 channels. We crop multiple entries by discarding sources which are closer than 2\farcs\, Finally, we perform a fidelity check to test the reliability of our candidates. Under the assumption that the positive emission comprehends real astrophysical sources and noise, while the negative emission are due only to noise, we use the following equation \citep{walter2016}:
\begin{equation}
    {\rm fidelity (SNR)} = 1 - \dfrac{N_{\rm neg} ({\rm SNR})}{N_{\rm pos} ({\rm SNR})} 
\end{equation}
with $N_{\rm neg} ({\rm SNR})$ and $N_{\rm pos} ({\rm SNR})$ are the number of negative and positive detections at a certain SNR, respectively. In order to not be limited by the low number counts of positive and negative detections around each quasar, we compute the fidelity considering the catalogs obtained from of all the quasars fields at once. Following \cite{venemans2020} and \cite{meyer2022b}, we assume a fidelity threshold as calculated in each channel of 90\% (Fidelity$>$0.9). We select sources with SNR$>$4, although it is worthwhile to notice that the cut on fidelity already selects sources with SNR$\gtrsim$4-5. We crop sources close to the edge of the field of view and those at a distance of $\Delta v > \pm 1000$ km s$^{-1}$\footnote{We note that changing this limit to $\Delta v > \pm 2000$ km s$^{-1}$ does not change our final resuls.} (e.g.\, \citealt{venemans2020}). 

Considering all the above conditions, we do not recover any companion candidate in our RL quasars fields. We can still compare this result with what observed around RQ quasars fields and with expectations from blank fields.

In particular, \cite{venemans2020} found 27 line emitters in the fields of 14 RQ quasars at $z\sim6$, out of 26 fields inspected. In particular, 19 (17) of these sources were identified as \cii\, emitters companion candidates to the quasars, with $\Delta v < 2000$ km s$^{-1}$ ($\Delta v < 1000$ km s$^{-1}$). They also discovered 3 companion candidates in the field of the RL quasar J2318-3113. However, our work and \cite{venemans2020} are characterized by different depths. If we consider solely the companion candidates around RQ quasars that would satisfy our SNR$>$4 criterion, for a \cii\, emission line with typical FWHM$\sim$300 km s$^{-1}$ ($F_{\rm \cii} \sim$1.77 Jy km s$^{-1}$), we are left with only 3 sources. These sources (in the fields of J0842+1218, PSO231-20, J2100-1715) were also recovered by \cite{decarli2017}, whose observations sensitivity is similar to that of our work. We also note that the companion candidates found by \cite{venemans2020} around J2338-3113 are fainter than the detection limits of our observations. Hence, at the depth of our work, one would expect to recover 3 \cii\, emitting companion candidates in 26 RQ quasars fields, i.e.\, in only $\sim$10\% of the fields. This is still consistent with finding zero companion candidates in 5 fields. 
We can also compare our results with the expected number of sources in a blank field, i.e.\, with no quasar. We consider the number density of \cii\, emitting sources provided in \cite{uzgil2021} and \cite{decarli2020}, in the redshift range $z=6-8$, with $L_{\rm \cii} > 2.8 \times 10^{8} L_{\odot}$\footnote{We note that our data is shallower, by a factor of $\sim$3, than those used to determine the blank field number density.} ($<$1.94$\times$10$^{-4}$ Mpc$^{-3}$; see Table 3 in \citealt{uzgil2021}). Considering the total volume covered in the 5 RL quasars fields observed here, within $\Delta v<1000$ km s$^{-1}$, we would expect to observe $<$0.02 sources. 

In conclusion, given the depth of our current observations and the number of fields explored, we are not able to place strong constraints on the properties of the fields of RL quasars, i.e.\, they are still consistent with both the expectations of blank fields and RQ quasars fields. Larger samples of RL quasars targeted in the sub-mm, and/or with deeper observations, will allow for further comparisons between the two samples.

\section{Discussion} \label{sec:disc}
The interaction between radio jets and the ISM of galaxies is expected to produce different outcomes, depending on various circumstances local to each galaxy and their proximate environment. As seen in both observations and simulations, a radio jet can drive massive, multi-phase (ionized, warm and molecular gas) outflows, whose geometry (following the jet direction, or perpendicular to it) depends on the jet inclination and jet-ISM coupling (e.g.\,\citealt{morganti2015}, \citealt{Meenakshi2022}, \citealt{Zhong2024}). This mechanism will also boost the gas turbulence, and hence increase the observed gas velocity and dispersion (e.g.\,\citealt{venturi2023}). In addition, the passage of the jet can form density fluctuations in the ISM, forming hot ($>150$ K) and dense ($>10^{5}$ cm$^{-3}$) gas, that can be observed via several CO emission lines, with an increase in the flux of higher J level transitions (e.g.\,\citealt{audibert2023}).
At the same time, observations of spectacular cases of RL AGN at low$-z$ (e.g.\,\citealt{guillard2015}, \citealt{appleton2018}) showed how the \cii\,emission line can be originated from shocks and/or outflows, and trace the radio jet-ISM interaction. \cite{pinchukova2019} reported a significant boost of \cii\, emission in a nearby AGN with radio jets, with a $L_{\rm \cii}/L_{\rm IR}$ ratio ten times higher than other AGN at comparable redshifts.
Considering all these, we might expect in RL quasars to observe \cii\,emission lines with larger FWHM, and/or asymmetric line profiles, or peculiar morphologies and with higher luminosities with respect to what recovered in RQ ones. On the contrary, the \cii\,luminosities of the sources in our sample are comparable or slightly lower than those observed in $z\sim6$ RQ quasars, and we do not recover any strong differences in FWHM values, spectral profiles or morphologies. This can be due to the fact that the aforementioned mechanisms might not be instantaneous, or could only be important on local scales, and hence difficult to highlight with our integrated, spatially unresolved, and low SNR observations (see also \citealt{molyneux2024}, for comparable conclusions in a study of a sample of $z<0.2$ radio AGN). Moreover, high spatial resolution observations highlight that several high$-z$ RQ quasars' hosts are found in mergers, which can broaden the apparent FWHM values \citep{neeleman2021}, complicating the comparison.
Another important aspect to consider is the possibility that extreme feedback from radio-jets can displace the gas and remove it altogether from the gravitational influence of the galaxy (e.g.\,\citealt{murthy2022}). The marginally lower \cii\,luminosities observed in our high$-z$ RL quasars sample may than be the result of such process. 
Finally, however, it is important to notice that the complex physical mechanisms responsible for \cii\,emission complicate the interpretation of our results. Higher SNR and angular resolution \cii\,observations are necessary to disentangle these different interpretations. 

Conversely, it is believed that radio-loud AGN are preferentially found in major mergers, which could trigger the nuclear accretion and the formation of the jets by funnelling a large amount of gas towards the center (e.g.\,\citealt{chiaberge2015}, \citealt{breiding2024}). At the same time, as discussed above, RL sources are expected to be found in overdensities of galaxies (e.g.\,\citealt{wylezalek2013}, \citealt{hatch2014}). \cite{khusanova2022} showed potential evidence for a larger merger fraction amongst $z>6$ RL quasars ($>40$\%) with respect to RQ ones ($\sim30$\%; \citealt{neeleman2021}). We do not recover merger signs in the RL quasars newly studied in this work. If we consider the entire sample of RL quasars (this work+literature), we recover only 2 sources which are potentially hosted in a major merger, allowing for a merger fraction of $\sim20$\%, even lower than that observed in RQ quasars. At the same time, none of the RL quasars inspected is surrounded by strong overdensities of bright \cii\,emitters. Given the course angular resolution and low SNR of our observations, and the limited sample reported, we refrain from drawing further conclusions. 

\section{Conclusions and Outlooks} \label{sec:conc}
In this work, we present new ALMA observations of six RL and one RQ quasars at $z>5$. We acquire data in Band 7 (5 RL sources), targeting the \cii\, emission line and underlying dust continuum, and in Band 3 (2 sources, 1 RL and 1 RQ), focusing on the \co\, line. We detect (at $\sigma>$2) the \cii\, emission line and continuum for the quasars targeted in Band 7 (Tables \ref{tab:1dcii}, \ref{tab:2dfit}), while no emission is recovered in Band 3, hence we can only place upper limits on the relative \co\, emission line flux/luminosity and underlying dust continuum/IR luminosity (Table \ref{tab:co65}).

At current spatial resolution ($\sim$1\farcs0$-$1\farcs4), no extended or disturbed morphologies are clearly detected in the Band 7 continuum maps (Figures \ref{fig:contmap}, \ref{fig:cont23map}), nor in the continuum-subtracted \cii\, emission line maps (Figure \ref{fig:1dciimap}): i.e.\, the source sizes are broadly consistent with the beam sizes. All the extracted \cii\, 1D spectra are well fitted with a single Gaussian function (Figure \ref{fig:1dciiline}). Hence, there is no strong evidence for mergers in the RL quasars hosts newly analyzed here.

Considering our new ALMA data and radio/sub-mm observations from the literature, we build the SED of the six RL quasars studied here (Figure \ref{fig:sed}). Although the observations forming the SEDs are obtained at different epochs and are not corrected for any potential variability, we observe that the ALMA 1mm measurements are consistent with being a combination of the high-frequency tail of radio synchrotron emission and dust continuum emission. At least in four sources this contribution is estimated to be $>9$\%, and in two cases, J1034+2033 and J0131-0321, it can be as high as $\sim$40\% and $\sim$100\%, respectively. Hence, extreme caution should be taken when deriving dust properties (luminosities, masses, SFR) for these objects.
Given the limited data at hands, a full spectral fit of the different components is currently out of the scope of this work.

We derive estimates on the \cii\, and infrared luminosities, as well as on the dust, dynamical and molecular gas masses, and \cii-\, and IR-based SFRs considering two scenarios: first, under the assumption that all the continuum emission arises from the galaxies' ISM, and secondly by accounting for our initial estimate of the synchrotron contribution (see Section \ref{sec:analysis_lum} and \ref{sec:analysis_massdyn}). The hosts of the RL quasars observed in this work present relatively large cool gas reservoirs, with $L_{\rm \cii} \sim 3-30 \times 10^{8}$, $M_{\rm dyn}\sim1-10 \times 10^{10} M_{\odot}$, and $M_{\rm H2, \cii}\sim9-80 \times 10^{9} M_{\odot}$. Moreover, they are also actively forming stars, with SFR$_{\rm \cii}\sim30-400$ $M_{\odot}$ yr$^{-1}$.

We place these new measurements in the context of observations of RQ and RL quasars' hosts from the literature. The \cii\, EW and FWHM in RL quasars are broadly consistent with those of RQ ones, but we observe slightly lower \cii\, luminosities and SFR for RL sources (with a p-value of $\sim$0.04; Figure \ref{fig:distrprop} and Table \ref{table:mean}).
Furthermore, the \cii\, decrement in RL quasars' hosts is in the same parameter space as those of RQ ones, with no dependency on radio loudness parameter. When considering the synchrotron contribution at $\sim$1mm, even outlier cases (e.g.\, J0131-0321 and J0410-0139) are consistent with RQ quasars (Figure \ref{fig:lfir_lciifir}).

Additionally, we search for line emitting candidates in the fields of the five RL quasars observed in Band 7 in this work, using the {\it findclumps} code, and selecting sources with a fidelity value higher than 0.9, a SNR$>4$, excluding the edges of the images, and within $\Delta v < 1000$ km s$^{-1}$ (see Section \ref{sec:analysis_comp}). We do not recover any \cii\,emitting companion candidate. Comparing this result with the number of sources expected in a blank field, and observed around a sample of RQ quasars at $z\sim$6, we see that the fields of RL quasars are still comparable with RQ ones or the blank field, given the depth of our data.

Several approaches will be important for further investigations of high$-$z RL quasars hosts properties. 
In the latest few years, the number of RL quasars at high$-z$ strongly increased (e.g.\,\citealt{gloudemans2022}, \citealt{belladitta2023}), allowing, in the future, to cover a larger sample of such sources and to undertake a more meaningful comparison with RQ quasars' hosts.  
In addition, radio/sub-mm observations of RL quasars over a large frequency range, covering the spectral region dominated by the jet and down to the peak of the modified black body dust emission, with several facilities (i.e., ALMA, VLA, NOEMA, uGMRT), will allow for a full fit of the spectral energy distributions. Hence, it will be possible to properly disentangle the non-thermal synchrotron and dust contributions, and to place strong constraints on the ISM properties of the host galaxies.
Moreover, follow-up observations of different sub-mm emission lines (e.g.\,[N\,II]\,205$\mu$m, [O\,I]\,146$\mu$m, [O\,III]\,88$\mu$m, [C\,I]\,369$\mu$m, high-J CO transitions) would enable constraints on the role of the central AGN and/or of the radio-jets in the arising of the \cii\, and continuum emission (e.g.\,\citealt{pensabene2021}, \citealt{li2024}). Additional higher spatial-resolution and SNR data targeted at the \cii\, emission line could allow to constrain more complex morphology/kinematic signatures, which might be due to merger events or the passage of a radio-jet (e.g.\,\citealt{neeleman2021}, \citealt{khusanova2022}).
The recently launched JWST already showed its key role in unveiling the properties of the hot ionized gas in both high-redshift quasars' hosts/galaxies (e.g.\, \citealt{ding2023}, \citealt{marshall2023}) and in powerful jetted/non-jetted AGN at Cosmic Noon (e.g.\, \citealt{wylezalek2022}, \citealt{cresci2023},\citealt{vayner2023}, \citealt{wangw2024}). Thus, deep NIRSpec IFU/NIRCAM observations of $z>5$ RL quasars' hosts would shed a fundamental light on the rest-frame optical properties of such galaxies, e.g.\,the underlying stellar continuum emission, the black hole mass and accretion rate from the H$\alpha$ or H$\beta$ emission line modelling, any central or extended outflow via observations of the [OIII] emission lines, and the characterization of the warm ionized ISM, and uncover any interplay with the radio-jets and their role in launching complex, multi-phase outflows. 

\begin{acknowledgements}
This paper makes use of the following ALMA data: ADS/JAO.ALMA\#2019.1.00840.S. ALMA is a partnership of ESO (representing its member states), NFS (USA) and NINS (Japan), together with NRC (Canada), MOST and ASIAA (Taiwan) and KASI (Republic of Korea), in cooperation with the Republic of Chile. The Joint ALMA Observatory is operated by ESO, AUI/NRAO and NAOJ.
This scientific work uses data obtained from Inyarrimanha Ilgari Bundara / the Murchison Radio-astronomy Observatory. We acknowledge the Wajarri Yamaji People as the Traditional Owners and native title holders of the Observatory site. CSIRO’s ASKAP radio telescope is part of the Australia Telescope National Facility (https://ror.org/05qajvd42). Operation of ASKAP is funded by the Australian Government with support from the National Collaborative Research Infrastructure Strategy. ASKAP uses the resources of the Pawsey Supercomputing Research Centre. Establishment of ASKAP, Inyarrimanha Ilgari Bundara, the CSIRO Murchison Radio-astronomy Observatory and the Pawsey Supercomputing Research Centre are initiatives of the Australian Government, with support from the Government of Western Australia and the Science and Industry Endowment Fund. This paper includes archived data obtained through the CSIRO ASKAP Science Data Archive, CASDA (https://data.csiro.au).
The National Radio Astronomy Observatory is a facility of the National Science Foundation operated under cooperative agreement by Associated Universities, Inc.
C.M. acknowledges support from Fondecyt Iniciacion grant 11240336 and the ANID BASAL project FB210003.
F.W. acknowledges support from the US NSF Grant AST-2308258.
R.A.M acknowledges support from the Swiss National Science Foundation (SNSF) through project grant 200020\_207349.
\end{acknowledgements}

%
%

%
%

\bibliography{aa_RLQSO_ALMA}
\bibliographystyle{aasjournal}

\begin{appendix} \label{app:velmapCII}
\section{Velocity maps}
\begin{figure*}
   \centering
   \includegraphics[width=0.75\hsize]{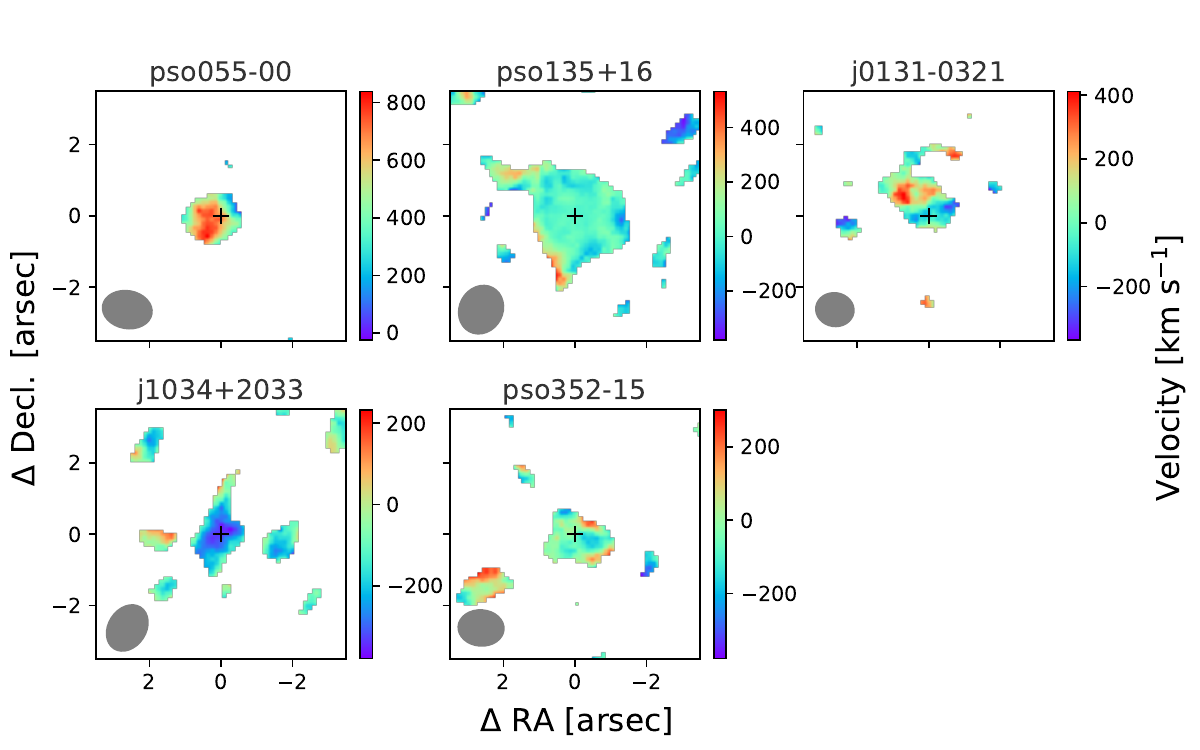}
   \caption{ALMA continuum-subtracted \cii\, velocity fields. }
              \label{fig:ciimapmom0}%
    \end{figure*}

We report here the continuum-subtracted \cii\, velocity maps of the RL quasars (see Figure \ref{fig:ciimapmom0}). We obtained these maps using the CASA task {\it immoments} with moments=1. We observe different kinematics: the majority of the quasars seem to be characterized by a uniform velocity fields, some with low velocities (e.g.\,PSO135+16, PSO352-15), another with velocities around $\sim -$200 km s$^{-1}$ (J1034+2033) and finally PSO055-00 characterized by very high velocities, up to $\sim$800 km s$^{-1}$. On the other hand, J0131-0321 seem to present more complex features, with a potential disk-like structure, with velocities between +400/$-$250 km $^{-1}$, and an elongated morphology. However, due to the low spatial resolution and limited SNR, more robust conclusions on the kinematical structure of our sources cannot be drawn. Further, higher resolution and SNR observations are necessary to firmly pin-point the velocity structures of high$-z$ RL quasars.

\end{appendix}

\end{document}